\documentclass[aps,prd,showpacs,a4paper,
 groupedaddress,preprintnumbers,10pt]{revtex4}
\usepackage{epsfig,graphicx,bm}
\begin{document}
\newcommand{\nd}{\noindent}
\newcommand{\beq}{\begin{equation}}
\newcommand{\eeq}{\end{equation}}
\newcommand{\barr}{\begin{eqnarray}}
\newcommand{\earr}{\end{eqnarray}}
\newcommand{\ba}{\begin{array}}
\newcommand{\ea}{\end{array}}
\newcommand{\bfk}{\mbox{\boldmath $k$}}
\newcommand{\be}{\begin{equation}}
\newcommand{\ee}{\end{equation}}
\newcommand{\bea}{\begin{eqnarray}}
\newcommand{\eea}{\end{eqnarray}}
\newcommand{\st}{{\scriptscriptstyle T}}
\newcommand{\xbj}{x_{\scriptscriptstyle B}}
\newcommand{\zh}{z_h}
\newcommand{\bfq}{\mbox{\boldmath $q$}}
\newcommand{\pup}{p^\uparrow}
\newcommand{\pdown}{p^\downarrow}
\newcommand{\qup}{q^\uparrow}
\newcommand{\qdown}{q^\downarrow}
\def\slash{\rlap{/}}
\newcommand{\bfp}{\mbox{\boldmath $p$}}
\newcommand{\bfP}{\mbox{\boldmath $P$}} 
\newcommand{\Lup}{\Lambda^\uparrow} 
\newcommand{\Ldown}{\Lambda^\downarrow} 
\newcommand{\Aup}{A^\uparrow} 
\newcommand{\hup}{h^\uparrow} 
\newcommand{\hdown}{h^\downarrow} 
\def\lsim{\mathrel{\rlap{\lower4pt\hbox{\hskip1pt$\sim$}}\raise1pt\hbox{$<$}}}
\def\gsim{\mathrel{\rlap{\lower4pt\hbox{\hskip1pt$\sim$}}\raise1pt\hbox{$>$}}}
\newcommand{\NP}[1]{ Nucl.\ Phys.\ {\bf #1}}
\newcommand{\ZP}[1]{ Z.\ Phys.\ {\bf #1}}
\newcommand{\PL}[1]{ Phys.\ Lett.\ {\bf #1}}
\newcommand{\PR}[1]{ Phys.\ Rev.\ {\bf #1}}
\newcommand{\PRL}[1]{ Phys.\ Rev.\ Lett.\ {\bf #1}}
\newcommand{\MPL}[1]{ Mod.\ Phys.\ Lett.\ {\bf #1}}
\newcommand{\SNP}[1]{ Sov.\ J.\ Nucl.\ Phys.\ {\bf #1}}
\newcommand{\EPJ}[1]{ Eur.\ Phys.\ J.\ {\bf #1}}
\newcommand{\IJMP}[1]{ Int.\ J.\ Mod.\ Phys.\ {\bf #1}}
\title{Parton intrinsic motion in inclusive particle production: \\
unpolarized cross sections, single spin asymmetries and the Sivers
effect}
\author{Umberto D'Alesio}
\email{umberto.dalesio@ca.infn.it}
\author{Francesco Murgia}
\email{francesco.murgia@ca.infn.it}
\affiliation{ Dipartimento di Fisica, Universit\`a di Cagliari and \\
Istituto Nazionale di Fisica Nucleare,  Sezione di Cagliari \\
 Casella Postale n. 170, I-09042 Monserrato (CA), Italy}
\vspace{8pt}

\date{\today}

\begin{abstract}
The relevance of intrinsic (or primordial) transverse momentum of
partons in the inclusive production of particles at high energy and 
moderately large $p_T$ has been known for a long time, beginning with
Drell-Yan and diphoton processes, and continuing with photon and meson 
production in hadronic collisions. In view of its renewed interest 
in the context of polarized processes and single spin asymmetries  
we perform, in the framework of perturbative QCD with the inclusion 
of spin and $\bm{k}_\perp$ effects, a detailed analysis
of several such processes in different kinematical
situations. We show that the inclusion of these effects leads, at the
level of accuracy reachable in this approach, to an overall satisfactory
agreement between theoretical predictions and experimental unpolarized
data, thus giving support to the study of spin effects and single spin
asymmetries within the same scheme. We present results for transverse
single spin asymmetries, generated by the so-called Sivers effect,
in inclusive pion and photon production in proton-proton
collisions. We compare our results with the available experimental
data and with previous results obtained using simplified
versions of this approach.
\end{abstract}
\maketitle
%
\section{Introduction}
\label{intr}

It has been known for a long time that the intrinsic (or primordial)
transverse momentum of partons inside hadrons involved in high-energy
processes may play a relevant role.  A typical example is the
Drell-Yan process, where the primordial, non-perturbative transverse
momentum of the initial partons is directly related to the lower
part of the transverse momentum spectrum of the observed lepton pair. 
A similar example is diphoton production in
hadronic collisions. Since the first applications of perturbative
quantum chromodynamics (pQCD) to inclusive particle production in
hadronic collisions, the role of these effects has been of some interest
and several generalizations of the usual collinear pQCD approach have
been presented \cite{fff} 
(throughout this paper, by collinear pQCD we mean the
approach in which intrinsic transverse momentum,
$\bm{k}_\perp$, effects are integrated
out up to a factorization scale and all partons/hadrons are assumed to
be collinear with parent hadrons/partons). In recent years, several
papers have reconsidered these effects, since collinear pQCD, even at
next-to-leading order (NLO) seems to underestimate experimental
results for photon and pion production in hadronic collisions in the
central rapidity and moderately large $p_T$ region \cite{pkt,ww,apa98}. 
It was found that the inclusion of intrinsic transverse momentum 
effects allows 
in most cases to reconcile theoretical calculations with experimental
results. This requires a relatively large average transverse momentum,
showing some dependence on the c.m.~energy of the process
considered. This probably indicates that some effects due to higher
order pQCD corrections are effectively embodied in the intrinsic
momentum contributions, a point which needs to be further clarified.

Almost independently of the above mentioned studies, the role of
intrinsic transverse momentum has received a lot of attention in the
context of polarization effects in inclusive particle production at
high energy and moderately large $p_T$, in particular regarding transverse
single spin asymmetries (SSA). Contrary to former expectations of
pQCD \cite{kane}, there are several experimental observations showing that
SSA can be large in this kinematical regime. It was originally suggested 
by Sivers \cite{siv} that pQCD with the inclusion of transverse momentum 
effects in parton distribution functions could be able to explain these 
results. Sivers' suggestion has been further extended in a number of 
subsequent papers \cite{noiS,newsiv}. There are several alternative possible
explanations for SSA, basically non-perturbative in nature. Recently,
Bourrely and Soffer \cite{bs03} have claimed that most of the
experimental data on SSA cannot be explained by pQCD, on the basis
that the collinear approach fails to reproduce the corresponding unpolarized
cross sections (which enter the denominator of the SSA) by 1-2 orders
of magnitude. As a matter of fact, in all previous papers on SSA
using pQCD-based approaches, a detailed analysis of unpolarized cross
sections was not performed. The main reason for this was that a number
of effects (scale dependences, NLO corrections) might cancel out in the SSA,
which are ratios of (polarized) cross sections.
Moreover, data on unpolarized cross
sections in the kinematical regions where SSA are measured to be
relevant are scarce. Considering the increasing theoretical and experimental 
interest in high-energy spin effects, a detailed combined analysis of 
unpolarized cross sections and SSA, within the same pQCD approach, is timely 
and worthwhile.

Based on these considerations, in this paper we first perform a
detailed analysis of a large amount of data on unpolarized cross
sections for inclusive particle production in several processes and in
different kinematical situations.  Let us stress that our approach and
aim are different in many respects from (and complementary to) those in
Ref.s \cite{pkt,ww,apa98}: our main interest remains the study of SSA.
Most of the interesting results on SSA are in kinematical regions
different from those preferentially considered in the above mentioned
papers, that is at 1 GeV$/c \lsim p_T \lsim 3$ GeV/$c$ and large $|x_F| = 
2|p_L|/\sqrt{s}$.  Moreover,
calculations of SSA are at present limited to the inclusion of 
leading-order contributions. Therefore, in this paper we limit
ourself to study unpolarized cross sections at the same level of
accuracy. In fact, our goal is not that of reproducing with very high 
precision the experimental results on unpolarized cross
sections. Rather, we want to show that the same approach used for SSA,
at the same level of accuracy, is able to reproduce unpolarized cross
sections for several processes and different kinematical situations,
up to a reasonable factor (of the order of 2-3) 
which may be attributed to NLO corrections
(the so-called $K$-factors), scale dependences, etc., which we expect,
at least, to partly cancel out in the SSA.

To this aim, we use here a generalization of the usual collinear
pQCD approach
to inclusive particle production with the inclusion of spin and
transverse momentum effects. We will take into account $\bm{k}_\perp$
effects for all partons involved in the process, including 
the fragmentation process where required, which
has been neglected in most of the Ref.s \cite{pkt,ww,apa98}.
This considerably improves the treatment of $\bm{k}_\perp$ effects
with respect to Ref.s \cite{noiS,noiC}, where a simplified approach
was adopted, including $\bm{k}_\perp$ effects only
at first non-vanishing order and with a simplified partonic
kinematics. Therefore, in this paper we will reconsider SSA for
several processes already analyzed in Ref.~\cite{noiS}, comparing the results
obtained in the two approaches.

The plan of the paper is the following: in section II we summarize our
approach, which will be largely utilized in the
rest of the paper. In section III we present and discuss our results
concerning the unpolarized cross sections for the Drell-Yan process
and for inclusive pion and photon production in hadronic collisions.  A
detailed comparison with available experimental results in several
kinematical configurations is given. In section IV we then present our
results for SSA in inclusive pion and photon production in hadronic
collisions, including only the so-called Sivers effect \cite{siv,noiS,newsiv}
as the possible mechanism for generating the SSA.
Our conclusions are presented in section V; the Appendix collects
useful relations regarding the kinematics.

\section{Phenomenological approach}
\label{appr}

In this section we present the basic ideas, ingredients and formulae of
our approach; details on kinematics are given in the Appendix.
Let us consider the inclusive production of large $p_T$ hadrons
in high-energy hadron collisions, that is the process
$AB\to C\,X$. The starting point of our approach is the
well-known collinear pQCD factorized expression for the
corresponding differential cross section:

\be 
\frac{E_C\,d\sigma^{AB\to C\,X}}{d^3\bm{p}_C} =
\sum_{a,b,c,d}\,\int dx_a\,dx_b\,dz\,f_{a/A}(x_a,Q^2)\,
f_{b/B}(x_b,Q^2)\,\frac{d\hat{\sigma}^{ab\to cd}}{d\hat{t}}\,
\frac{\hat s}{\pi z^2}\,\delta(\hat{s}+\hat{t}+\hat{u})\,D_{C/c}(z,Q^2)\,.
\label{cpqcd}
\ee

Here $x_a$, $x_b$ ($z$) are the light-cone momentum fractions of
partons $a$, $b$ (hadron $C$) with respect to the parent hadrons $A$,
$B$ (parton $c$); $f(x,Q^2)$, $D(z,Q^2)$ are the well-known
non-perturbative parton distribution functions (PDF) and fragmentation
functions (FF) respectively, where the intrinsic transverse momentum
dependence has been integrated over up to the factorization scale
$Q^2$. Furthermore, $\hat{s}$, $\hat{t}$, $\hat{u}$ are the Mandelstam 
variables for the partonic scattering process, $ab\to cd$, with all
(massless) partons taken to be collinear with the corresponding
hadrons. The Dirac delta function accounts for energy-momentum
conservation in the elementary
process and allows to express e.g.~$z$ as a function of $x_a$ and
$x_b$ and the {\it hadronic} Mandelstam invariants (see also the
Appendix).

Extending this formalism with the inclusion of intrinsic transverse
momentum of partons is not trivial at all and poses several
problems. To start with, a complete formal proof of the factorization
theorem is still missing in this case. The validity of factorization
has been conjectured by Collins \cite{col} and proved only for the
Drell-Yan process and in $e^+e^-$ annihilation \cite{css}. Quite
recently a factorization formula for semi-inclusive deep inelastic
scattering in the current fragmentation region has been proved
\cite{jimay}. 
A formally complete definition of transverse-momentum dependent PDF's
and FF's and of their evolution properties is still missing. In a partonic
approach and at leading twist in the factorization scale $Q$ the
usual collinear parton densities $f_{a/A}(x_a,Q^2)$ are simply
generalized to transverse-momentum dependent distributions
$\hat{f}_{a/A}(x_a,\bm{k}_{\perp a},Q^2)$, where $\bm{k}_{\perp a}$ 
is the transverse
momentum of parton $a$ with respect to the light-cone direction of
hadron $A$, such that
\be 
f_{a/A}(x_a,Q^2) = \int d^2\bfk_{\perp a} \, \hat
f_{a/A}(x_a,\bfk_{\perp a},Q^2)
\>.
\label{fxk}
\ee
Analogously, the usual parton fragmentation function $D_{C/c}(z,Q^2)$
generalizes to $\hat{D}_{C/c}(z,\bm{k}_{\perp C},Q^2)$, where
$\bm{k}_{\perp C}$ is
the transverse momentum of the final hadron $C$ with respect to the
light-cone direction of the fragmenting parton $c$, and
\be 
D_{C/c}(z,Q^2)=\int
d^2\bm{k}_{\perp C}\,\hat{D}_{C/c}(z,\bm{k}_{\perp C},Q^2)\,.
\label{dxk}
\ee
Intrinsic transverse momentum effects are of higher twist; consistently,
a complete higher-twist treatment of the process would be
required. However, we are far from being able to perform this complete
analysis, since it introduces new, unknown non-perturbative PDF's and
FF's and quark-gluon correlations; furthermore, the usual partonic
interpretation of PDF's and FF's would be lost in this case.

Despite these problems, there is a strong  phenomenological evidence
that transverse momentum effects in the partonic kinematics and
dynamics are very relevant and may contribute to account for several 
experimental results for (un)polarized cross sections in inclusive particle 
production, difficult to explain in the collinear
pQCD approach.

This is why this generalized approach has been extensively employed in 
recent years; some of the resulting phenomenological outcomes are quite
promising, although much work remains to be done on its formal
aspects.  This leads to a modified expression for
the differential cross section, Eq.~(\ref{cpqcd}), which now reads
\begin{eqnarray}
\frac{E_C\,d\sigma^{AB\to C\,X}}{d^3\bm{p}_C} &=& \sum_{a,b,c,d}\,\int
dx_a d^2\bm{k}_{\perp a}\,dx_b d^2\bm{k}_{\perp b}\,dz d^3\bm{k}_C\,
\delta(\bm{k}_C\cdot \hat{\bm{p}}_c)\,\hat{f}_{a/A}(x_a,\bm{k}_{\perp a},Q^2)\,
\hat{f}_{b/B}(x_b,\bm{k}_{\perp b},Q^2) \nonumber\\
&\times&\>\frac{\hat s}{x_ax_b s}
\frac{d\hat{\sigma}^{ab\to cd}}{d\hat{t}}(x_a,x_b,\hat{s},\hat{t},
\hat{u})\,\frac{\hat s}{\pi}\,\delta(\hat{s}+\hat{t}+\hat{u})\,
\frac{1}{z^2}\,J(z,|\bfk_C|)\,\hat{D}_{C/c}(z,\bm{k}_C,Q^2)\,.
\label{kqcd}
\end{eqnarray}

Notice that the partonic cross sections $d\hat{\sigma}/d\hat{t}$
depend, via the Mandelstam variables, on the intrinsic transverse
momenta of partons.  Notice also that the integration over the
transverse momentum of the observed hadron $C$ with respect to the
light-cone direction of the
fragmenting parton $c$ runs over a generic three-momentum $\bm{k}_C$,
defined in the c.m. of the initial hadrons $A$ and $B$ (we will call
this the {\it hadronic} c.m. frame in the following); the
delta function $\delta(\bm{k}_C\cdot \hat{\bm{p}}_c)$ guarantees that
$\bm{k}_C$ is in fact always orthogonal to $\bm{p}_c$, the parton $c$
three-momentum, and also the correct normalization of the
fragmentation function, according to Eq.~(\ref{dxk}), which makes
$\bm{k}_C$ completely equivalent to $\bm{k}_{\perp C}$.
The extra phase-space factor $J(z,|\bfk_C|)/z^2$ is the proper invariant 
Jacobian factor for the transformation from the parton momentum
$\bfp_c$ to the hadron
momentum $\bfp_C$ with inclusion of transverse momentum effects; 
the term $\hat s/(x_a x_b s)$ restores proper 
flux factor for non-collinear (in the hadronic c.m. frame)
parton-parton collision (see Appendix).

Let us stress at this point that, once collinear pQCD is
complemented with the inclusion of $\bm{k}_\perp$ effects, not only
Eq.~(\ref{cpqcd}) modifies to Eq.~(\ref{kqcd}), but additional
contributions to the unpolarized cross section are in principle
possible \cite{muld,boer,abdlmm}.
These contributions to the unpolarized cross section
could play a relevant role when considering asymmetries with respect to
some measurable azimuthal angle related to the partonic process
(e.g. in the Drell-Yan process, see \cite{boer}); however, we have 
explicitly checked that they give negligible contributions to the 
cross section for the inclusive process $AB\to C\,X$, where all
``internal'', partonic variables are integrated over. Therefore, they
are of little relevance for this analysis,
and will not be further considered in the rest of the paper. 
A more complete discussion will be given elsewhere \cite{abdlmm}.

Eq.~(\ref{kqcd}) has been widely used in the literature, starting from
the pioneering work of Feynman, Field and Fox~\cite{fff}.
Several papers have shown in the recent past that intrinsic
$\bfk_\perp$'s are indeed necessary in order to improve the agreement
between next-to-leading order (NLO) pQCD calculations and experimental
data for inclusive pion and photon production at high energies and 
moderately large $p_T$ \cite{pkt,ww,apa98}.

An expression similar to Eq.~(\ref{kqcd}) holds also for the numerator of
a SSA, $(d\sigma^\uparrow-d\sigma^\downarrow)/
(d\sigma^\uparrow+d\sigma^\downarrow)
\propto d\Delta^{\!N}\sigma/d\sigma$,
replacing, for the polarized particles involved, the corresponding 
unpolarized PDF, FF and partonic cross sections by the appropriate 
polarized ones, $\Delta^{\!N}\!f$, $\Delta^{\!N}\!D$,
$d\Delta^{\!N}\hat{\sigma}$ (see Ref.~\cite{noiC} and section
\ref{tssa} for more details).
At leading twist there are four such new spin and
$\bfk_\perp$ dependent functions to take into account:
\begin{eqnarray}
\Delta^{\!N}\!f_{a/A^\uparrow} &\equiv& \hat f_{a/A^\uparrow}(x,
\bfk_{\perp})-\hat f_{a/A^\downarrow} (x, \bfk_{\perp})\>;\>\>\>\>\>\>
\Delta^{\!N}\!f_{a^\uparrow/A} \;\equiv\; \hat f_{a^\uparrow/A}(x,
\bfk_{\perp})-\hat f_{a^\downarrow/A} (x, \bfk_{\perp})\>;
\label{delf2} \\ 
\Delta^{\!N}\!D_{C/c^\uparrow} &\equiv & \hat
D_{C/c^\uparrow}(z, \bfk_{\perp}) - \hat D_{C/c^\downarrow}(z,
\bfk_{\perp})\>; \>\> \Delta^{\!N}\!D_{C^\uparrow/c} \;\equiv\; \hat
D_{C^\uparrow/c}(z, \bfk_{\perp}) - \hat D_{C^\downarrow/c}(z,
\bfk_{\perp})\,,
\label{deld2}
\end{eqnarray}
\noindent two in the PDF sector, Eq. (\ref{delf2}), and two in the FF
sector, Eq. (\ref{deld2}); the first functions in Eq.s (\ref{delf2}),
(\ref{deld2}) are respectively the so-called Sivers \cite{siv} and
Collins  \cite{col} function.  The second ones are respectively the
function introduced by Boer and Mulders \cite{muld,boer} and the so-called
``polarizing'' FF \cite{muld,abdm1}. 
For a general overview on spin and $\bfk_\perp$-dependent PDF's and
FF's see also Ref.~\cite{enzo}.

Regarding SSA, in this paper we will focus on the Sivers effect in
inclusive pion and photon production.
For a phenomenological study of the role of the Collins
effect within the same approach and at the same level of accuracy see
\cite{abdlm}. SSA and Sivers effect in the Drell-Yan process 
have been discussed in \cite{adm1}.

In the following sections, we will first concentrate on unpolarized
cross sections, applying Eq.~(\ref{kqcd}) to several processes in
different kinematical situations. Some details of the approach
may differ for different processes and will therefore be discussed in the
appropriate sections. We now briefly comment on how some more general
ingredients required in practical calculations based on
Eq.~(\ref{kqcd}) have been fixed.

\noindent {\bf 1)}
Concerning the $\bm{k}_\perp$-dependent PDF's (FF's) we assume that
the $x$ ($z$) and $\bm{k}_\perp$ dependences factorize; we use a
Gaussian-like, flavour-independent behaviour for the
$\bm{k}_\perp$-dependent part;
that is, neglecting for the moment the dependence on the factorization
scale, we take ($k_\perp=|\bm{k}_\perp|$)
\be 
\hat f_{a/A}(x,\bfk_{\perp a}) = f_{a/A}(x)\,\frac{\beta^2}{\pi}\>
e^{-\beta^2\,k_{\perp a}^{\,2}}\>; \qquad\qquad \hat
D_{C/c}(z,\bfk_{\perp C}) = D_{C/c}(z)\,\frac{\beta'^2}{\pi}\>
e^{-\beta'^2\,k_{\perp C}^{\,2}}\>,
\label{gk}
\ee
\noindent where the parameter $\beta$ ($\beta'$) is related to the
average partonic (hadronic) $k_{\perp}$ by the simple relation
$1/\beta(\beta')=\langle\,k_{\perp a(C)}^2\,\rangle^{1/2}$.

\noindent {\bf 2)}
The cross sections for the elementary, partonic scattering,
$d\hat{\sigma}^{ab\to cd}/d\hat{t}$, are calculated at leading order
in the strong coupling constant power expansion,
including in the partonic kinematics the full dependence on the
intrinsic transverse momenta $\bfk_\perp$. That is, the partonic cross
sections in Eq.~(\ref{kqcd}) will depend on the properly
$\bm{k}_\perp$-modified partonic invariants $\hat s$, $\hat t$
and $\hat u$. At relatively low $p_T$ the inclusion of $\bm{k}_\perp$
dependence might make one or more of the partonic Mandelstam variables 
become smaller than a typical hadronic scale.
This configuration would correspond to a
situation where the propagator of the exchanged particle in the
partonic scattering becomes soft; the observed transverse momentum is
thus generated mainly by fluctuations in the intrinsic
$\bm{k}_\perp$ distribution and not by the hard scattering. In this case
perturbation theory would break down. In order to avoid such a problem
and extend this approach down to $p_T$ around 1-2 GeV$/c$ (where most of
data on single spin asymmetries have been collected) different ways
have been proposed.  Following \cite{pkt}, we introduce a regulator
mass, $\mu=0.8$ GeV, shifting all partonic Mandelstam variables, that
is we take 
\be 
\hat t \to \hat t -\mu^2,\;\;\;\; \hat u \to \hat u
-\mu^2,\;\;\;\; \hat s \to \hat s + 2\mu^2\>.  
\ee

\noindent {\bf 3)}
Another somehow related source of potential ambiguity is the behaviour of the
strong coupling constant, $\alpha_s(Q^2)$, in the low $Q^2$ regime.  We
adopt the prescription originally proposed by Shirkov and Solovtsov
\cite{ss}, using for $\alpha_s$ the expression
\be
\label{ass}
\alpha_s(Q^2) = \frac{1}{\beta_0}\left[ \frac{1}{\log (Q^2/\Lambda^2)} +
\frac{\Lambda^2}{\Lambda^2-Q^2} \right] \,,
\ee 
where as usual $\beta_0 =
(33-2n_f)/12\pi$, $n_f$ being the number of active flavours (we use
$n_f=4$), and $\Lambda=0.2$ GeV/$c$.  According to
Eq.~(\ref{ass}), at large $Q^2$ $\alpha_s$ reduces to the standard LO
expression, while at low $Q^2$ its behaviour is well
under control without the introduction of any extra parameter, like a 
freezing scale parameter or a dynamical gluon mass. We have
explicitly checked that other prescriptions give similar results.

Note that for the Drell-Yan process Eq.~(\ref{kqcd}) simplifies a lot
(see section \ref{unpdy}); moreover, the large scale involved (namely
the invariant mass of the observed lepton pair) removes the danger of 
critical kinematical regions discussed above.

\noindent {\bf 4)}
Another important ingredient is the choice of the factorization scale
$Q$ (we use one single scale for renormalization and factorization
scales) governing the pQCD evolution of PDF's and FF's and entering
the QCD running coupling constant, $\alpha_s$.
It is well known that LO calculations,
and in some cases NLO too, can strongly depend on this choice.
Moreover, there is not in the literature a unique prescription for 
the choice of the factorization scale $Q$ in inclusive hadron production, and
several possible alternatives have been suggested. 
Typical scales adopted are the observed hadron transverse momentum,
$p_T$, or $p_T/2$, and the transverse momentum of the fragmenting
parton in the partonic c.m frame, $\hat{p}^*_T$, or $\hat{p}^*_T/2$.
Apart from the Drell-Yan process, where the natural scale is the 
invariant mass of the produced lepton pair, throughout this paper 
we will adopt the factorization scale $Q=\hat{p}^*_T/2$.

\noindent {\bf 5)}
Theoretical results for cross sections at LO and NLO
are usually compared to get an estimate of the 
uncertainty related to the choice of the factorization scale 
and indications on the convergence 
of the perturbative expansion. The ratio of cross sections
evaluated at NLO and LO respectively gives the so-called
$K$-factor for the process under consideration. It is often assumed \cite{pkt}
that this ratio is independent of intrinsic transverse momentum
effects, so that it can be estimated in the usual collinear pQCD
approach.  Throughout this paper we will conform to this prescription;
that is, in comparing our LO results for cross sections with the 
corresponding experimental results, we will include collinear pQCD 
$K$-factors evaluated by means of independent numerical codes
\cite{aue,ave,wer}.
For the processes under consideration in this paper, the $K$-factors 
show some general behaviour with respect
to the relevant kinematical variables~\cite{wer}:
they decrease smoothly with the increasing of the observed hadron
transverse momentum, $p_T$ (the variation being steeper at the lower 
$p_T$ considered), at fixed $\sqrt{s}$; they also decrease mildly with
$\sqrt{s}$ at fixed $p_T$, whereas they show a tiny dependence on $x_F$.
For simplicity and clarity, in our plots of cross sections we will
always show our numerical curves rescaled by a fixed (for that curve) 
$K$-factor, obeying the general trends of above and chosen according
to the results of the numerical code INCNLL of Ref.s \cite{aue,ave}.
We think this should help 
in clarifying the role played by $K$-factors in our results. The mild 
variations of the computed $K$-factor for each point of a given curve
should not change our general conclusions. More details on the
estimate of $K$-factors are given in the following sub-sections.

Let us finally recall that our main aim is not to perform 
an overall best fit to the available data (to this end, a full NLO
approach would be required). Rather, we want to show to what extent the same 
approach adopted in the description of SSA is able, when complemented 
by collinear pQCD $K$-factors, to give a reasonably good account of a 
large set of cross-section data for different processes in different 
kinematical situations.  We feel that the approach here described is 
reliable in this respect.
Notice also that, as we have explicitly checked,
the main interest of our study, namely single spin asymmetries, 
are not strongly affected by the choices discussed above.

In the following sub-sections we discuss separately and in more detail 
the processes under study.

\subsection{Drell-Yan process}
\label{unpdy}
Let us now consider lepton pair production
in $pp$, $p\bar{p}$ collisions, the Drell-Yan process.  In LO, collinear
pQCD the lepton pair has no transverse momentum, $\bfq_T$, in
the hadronic center of mass frame.  On the other hand plenty of data
show a well-established $q_T$ spectrum, with an exponential
behaviour at low $q_T$ ($\alt 2-3$ GeV$/c$) turning into a power-like
one at larger transverse momentum.
NLO corrections and soft gluon radiation are important
contributions both for the normalization of the cross section and in
order to reproduce the observed power-like behaviour at large $q_T$.  
However, to describe the behaviour at very small $q_T$ one has to 
include the intrinsic parton momentum, $\bfk_\perp$.

Therefore, within our approach the Drell-Yan process can be considered a
useful tool to gain information on the intrinsic transverse momentum
of partons inside the initial colliding hadrons.
 
Analogously to Eq.~(\ref{kqcd}), the cross section for the production
of an $\ell^+\ell^-$ pair in the collision of two hadrons $A$ and $B$
(there is no need for any fragmentation function here) reads:
\be 
d\sigma = \sum_{ab} \int  dx_a \, d^2\bfk_{\perp a} \, dx_b
\, d^2\bfk_{\perp b} \, \hat f_{a/A}(x_a,\bfk_{\perp a}) \,
\hat f_{b/B}(x_b,\bfk_{\perp b}) \,
\frac{\hat{s}}{x_a x_b s}\,
 d\hat\sigma^{ab \to \ell^+\ell^-}\,,
\label{dy1}
\ee
where the elementary cross section $d\hat\sigma$ for the process
$a(p_a) \, b(p_b) \to \ell^+(p_+) \, \ell^-(p_-)$ is given by:
\be 
d\hat\sigma = \frac{1}{2\hat s} \> \frac{d^3p_+}{2E_+} \>
\frac{d^3p_-}{2E_-} \> \frac{1}{(2\pi)^2} \> \delta^4(p_a + p_b - p_+
- p_-) \> \overline{\left\vert \, M_{ab \to \ell^+\ell^-} \,
\right\vert^2} \>.
\label{ecs}
\ee

The differential cross section $d\sigma$ depends on the variables
\be 
\hat s \equiv M^2 = (p_a + p_b)^2 \equiv q^2\,, \quad\quad\quad y
= \frac 12 \ln \frac{q_0 + q_L}{q_0 - q_L} \,,\quad \quad\quad \bfq_T
\>,
\label{var}
\ee
that is the squared invariant mass, the rapidity and the transverse
momentum of the lepton pair; $q_0$, $\bfq_T$ and $q_L$ are
respectively the energy, transverse and longitudinal components, in
the $A$-$B$ c.m. frame, of the four-vector $q = p_a + p_b =
p_+ + p_-$. Using the relations:
\be 
\frac{d^3p_-}{2E_-} = d^4p_- \, \delta(p_-^2) \quad\quad\quad p_-
= q - p_+ \quad\quad\quad dM^2 \, dy = 2 \, dq_0 \, dq_L \>,
\label{jacdy}
\ee
Eq.~(\ref{dy1}) can be written as
\be 
\frac{d^4\sigma}{dy \, dM^2 \, d^2\bfq_T} = \sum_{ab}\! \int\!
 dx_a \, d^2\bfk_{\perp a} \, dx_b \, d^2\bfk_{\perp b} 
\hat f_{a/A}(x_a,\bfk_{\perp a}) \, \hat f_{b/B}(x_b,\bfk_{\perp b})
\, \delta^4(p_a + p_b - q) \,
\frac{\hat{s}}{x_a x_b s}\,
 \hat\sigma^{ab}_0 \,,
\label{ddy2} 
\ee
where $\hat\sigma^{ab}_0$ is the total cross section for the $ab \to
\ell^+\ell^-$ process:
\be 
\hat\sigma_0^{ab} = \int \frac{d^3p_+}{2E_+} \> \frac{1}{(2\pi)^2}
\frac{1}{2M^2} \> \delta((q-p_+)^2) \> \overline{\left\vert \, M_{ab
\to \ell^+\ell^-} (p_+,q) \, \right\vert^2} \>. 
\label{ecs0} 
\ee

For the kinematical regimes of interest in this paper the dominating
elementary contribution to the Drell-Yan process is the lowest order
electromagnetic interaction, $q \bar q \to \gamma^* \to \ell^+\ell^-$,
so that $a,b = q,\bar q$ with $q = u, \bar u, d, \bar d, s, \bar s$
and
\be 
\hat\sigma^{q\bar q}_0 = \frac{4\,\pi \, \alpha^2 \,
e_q^2}{9\,M^2} \>\cdot
\label{s0}
\ee
  
The Dirac-$\delta$ function accounting for energy-momentum conservation
in Eq.~(\ref{ddy2}) contains the factors 
\bea 
&&\frac{1}{2} \, \delta(E_a + E_b - q_0) \, \delta(p_{za} +
p_{zb} - q_L) = \nonumber \\ 
&&\frac{1}{2} \, \delta \! \left( (x_a + x_b)\frac{\sqrt s}{2} + 
\left[ \frac{k_{\perp a}^2}{x_a s} +
\frac{k_{\perp b}^2}{x_b s} \right] \frac{\sqrt s}{2} - q_0 \right)
\times \nonumber \\ 
&& \delta \! \left( (x_a - x_b)\frac{\sqrt s}{2} -
\left[ \frac{k_{\perp a}^2}{x_a s} - \frac{k_{\perp b}^2}{x_b s}
\right] \frac{\sqrt s}{2} - q_L \right) \>. 
\label{ndel} 
\eea

In the following we shall only consider kinematical regions such that:
\be 
q_T^2 \ll M^2 \quad\quad\quad k_{\perp a,b}^2 \simeq q_T^2 \>, 
\ee
where Eq.~(\ref{ndel}) simplifies into the usual collinear condition:
\be 
\frac{1}{2} \, \delta(E_a + E_b - q_0) \, \delta(p_{za} + p_{zb} -
q_L) = \frac{1}{s} \, \delta \! \left( x_a - \frac{M}{\sqrt s} \, e^y
\right) \, \delta \! \left( x_b - \frac{M}{\sqrt s} \, e^{-y} \right)
\>. 
\label{odel} 
\ee

The Gaussian shape adopted for the $\bfk_\perp$ dependent PDFs',
Eq.~(\ref{gk}), together with Eq.~(\ref{odel}), allow us to
analytically perform the integrations in Eq.~(\ref{ddy2}) which becomes
\be 
\frac{d^4\sigma}{dy\,dM^2\,d^2\bfq_{T}} =
\frac{\hat \sigma_0}{\pi s} \, \frac{\beta^2 \bar{\beta}^2}{\beta^2 +
\bar{\beta}^2}\exp \left[ - \frac{\beta^2\bar{\beta}^2} {\beta^2 +
\bar{\beta}^2} \, q_{T}^2 \right] \> \sum_q e_q^2 \, f_{q/p}(x_a) \,
f_{\bar q/p}(x_b) \label{dyfin}\>,  
\ee

\noindent
where $\beta$, $\bar{\beta}$ refer to the quark(antiquark) PDF
respectively.  By direct comparison with data we can then extract the
$\beta$, $\bar{\beta}$ parameters. The numerical values obtained in
this way are of course related to the set of $x$ dependent PDF's
adopted; throughout this paper we use the MRST01 set \cite{mrst01}.
In principle $\beta$ and $\bar{\beta}$ could be $x$ and flavour
dependent, but in first approximation a reasonable description of data
can be obtained by neglecting these dependences.
In this case, $\bar{\beta}=\beta$ and $1/\beta^2=\langle q_T^2 \rangle/2$.

There is a clear experimental evidence that the average transverse
momentum $\langle q_T\rangle$ of the lepton pair increases with the
c.m. energy, in agreement with pQCD calculations \cite{app}.  In our LO
approach, see Eq.~(\ref{dyfin}), this behaviour can be obtained by
using an effective value of $\beta$ decreasing with 
c.m. energy. This way, one should more correctly interpret the
parameter $\beta$ as representing the effects of the {\em primordial}
transverse motion of partons plus a component coming from NLO corrections.
Since our first aim here is to extract information on the
primordial intrinsic momentum, we write 
\be 
\langle q_T^2 \rangle=\langle q_T^2 \rangle_{\rm{intr}}
+\langle q_T^2 \rangle_{\rm{pert}}\,, 
\ee 
where $\langle q_T^2 \rangle_{\rm{pert}}\propto \alpha_s\, s$, is a
perturbative, energy dependent contribution, which will be estimated
by comparison with data. A good description of
data available in different kinematical regions can be obtained by
choosing at the lowest c.m. energies considered, $\sqrt{s} \simeq 20$ GeV, 
\be
(\,\langle q_T^2\rangle_{\rm{intr}}/2\,)^{1/2}=
1/\beta_0 = 0.8 \;\; {\rm GeV}/c\>.
\label{beta}
\ee

\begin{figure} [!,h,t]
\epsfig{figure=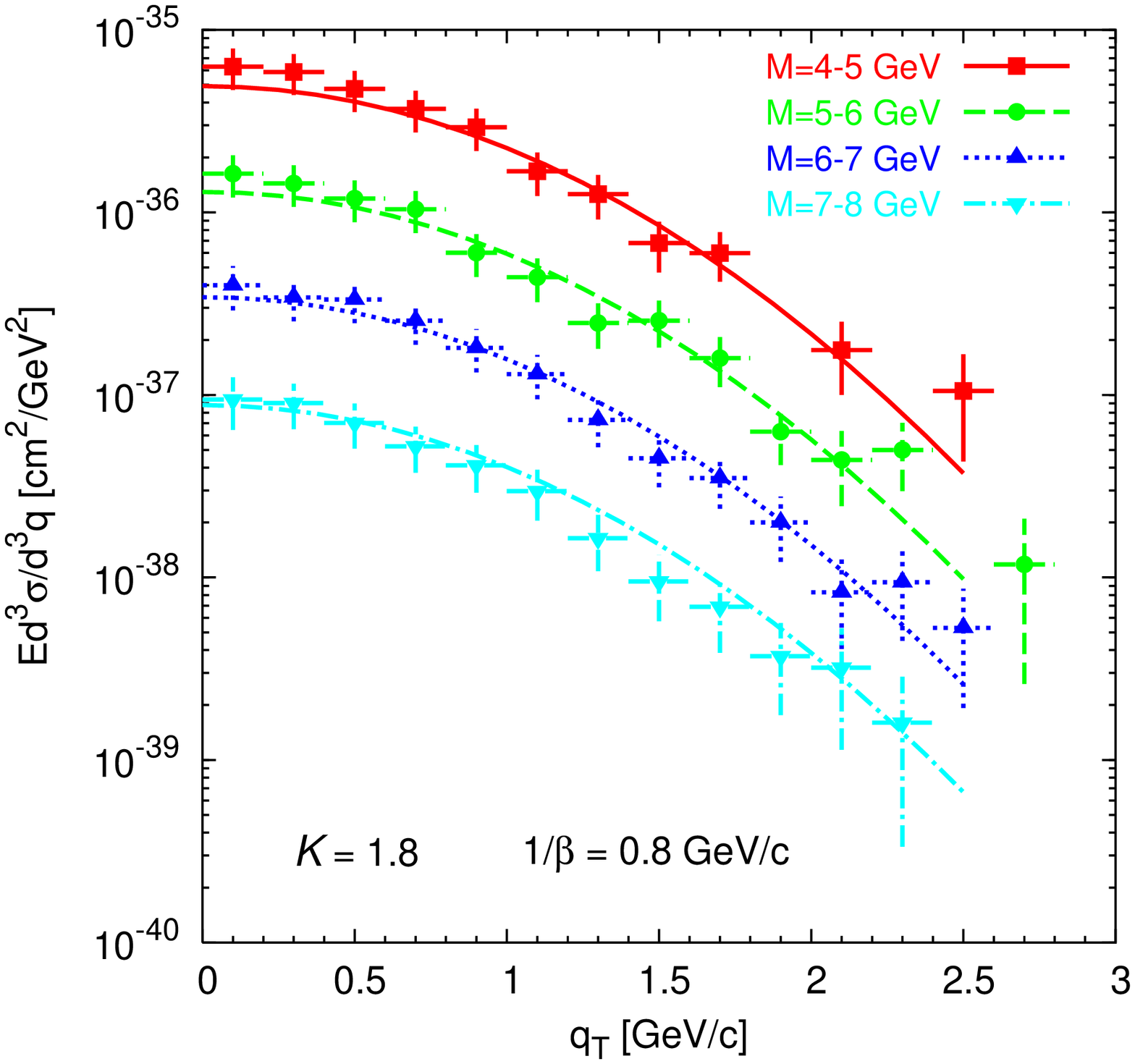,angle=-90,width=.63\textwidth}
\caption{ Invariant differential cross section for the Drell-Yan
process at $\sqrt{s}\simeq 19.4$ GeV and fixed rapidity $y=0.4$, as a
function of the transverse momentum of the lepton pair $q_T$ and
averaged over different invariant mass bins (see the legend). The
parameterization MRST01 \cite{mrst01} for the unpolarized parton
distributions is used, with $1/\beta = 0.8$ GeV$/c$.  
Curves are rescaled by a fixed $K$-factor, $K=1.8$. 
Data are from \cite{ito}.}
\label{dy200}
\end{figure}

\begin{figure} [!,h,b]
\epsfig{figure=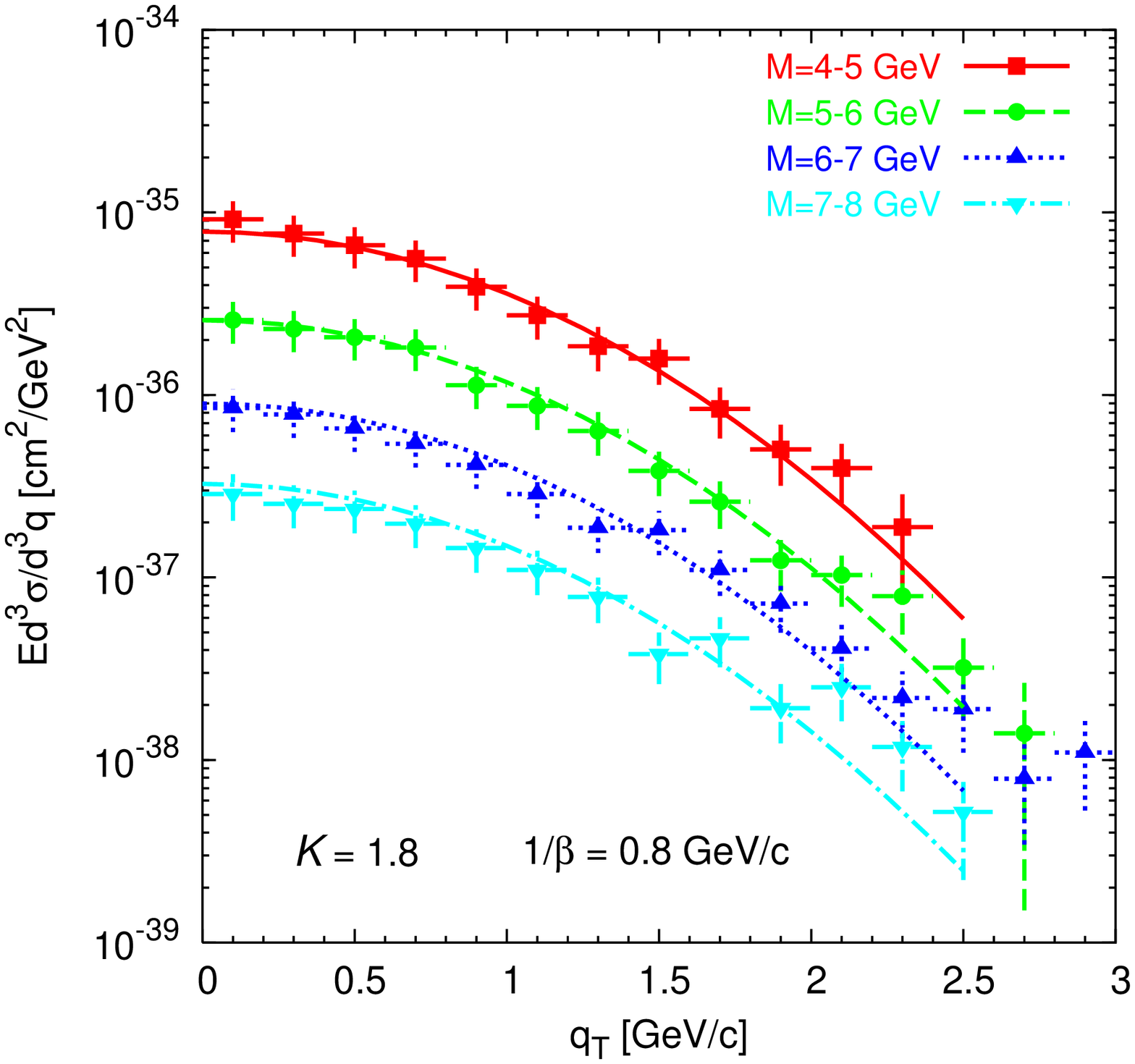,angle=-90,width=.63\textwidth}
\caption{ Invariant differential cross section for the Drell-Yan
process at $\sqrt{s}\simeq 23.8$ GeV and fixed rapidity $y=0.21$, as
a function of the transverse momentum of the lepton pair $q_T$ and
averaged over different invariant mass bins (see the legend). The
parameterization MRST01 \cite{mrst01} for the unpolarized parton
distributions is used, with $1/\beta = 0.8$ GeV$/c$. 
Curves are rescaled by a fixed $K$-factor, $K=1.8$. 
Data are from \cite{ito}.}
\label{dy300}
\end{figure}

\begin{figure} [!,h,t]
\epsfig{figure=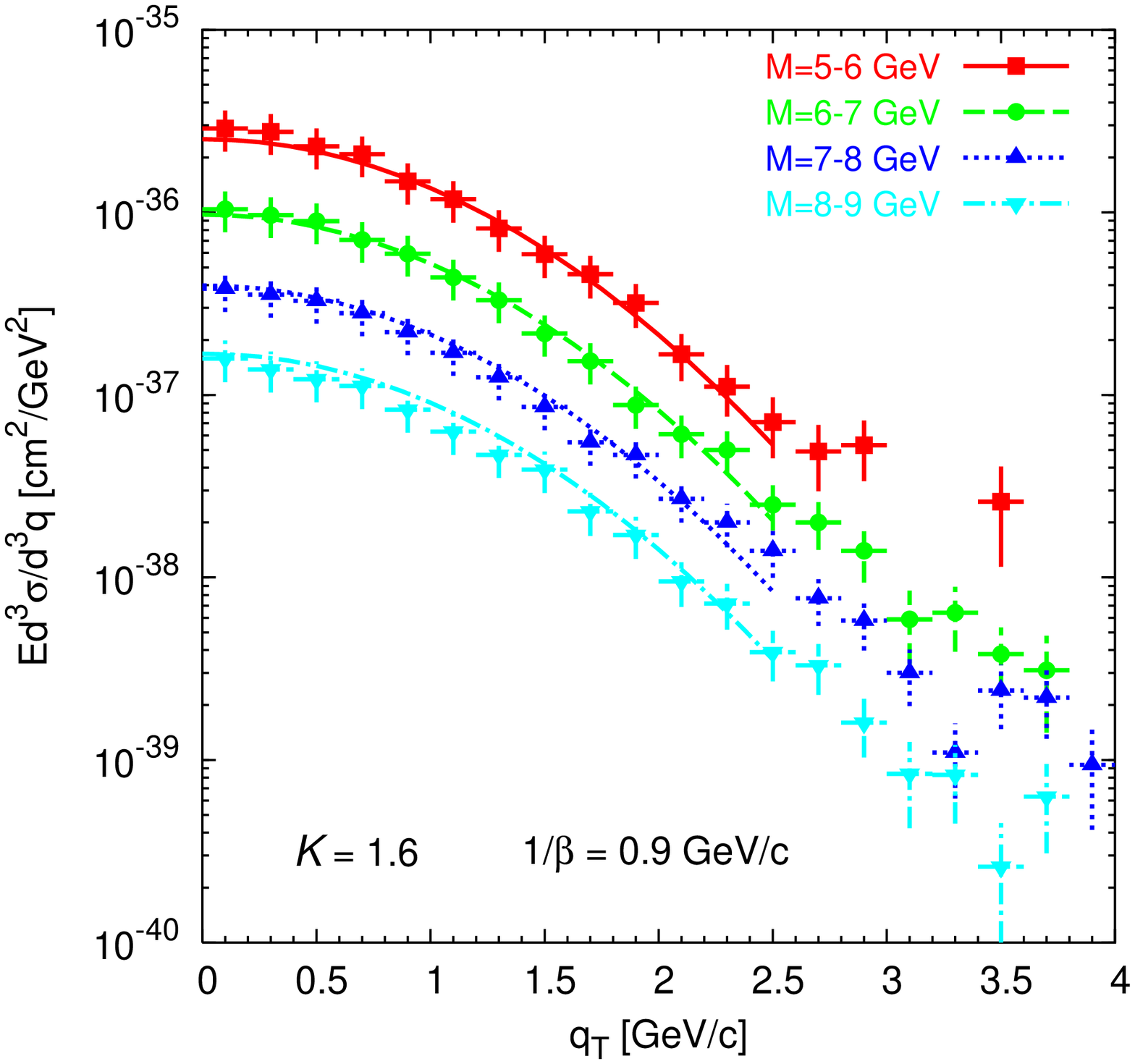,angle=-90,width=.63\textwidth}
\caption{ Invariant differential cross section for the Drell-Yan
process at $\sqrt{s}\simeq 27.4$ GeV and fixed rapidity $y=0.03$, as a
function of the transverse momentum of the lepton pair $q_T$ and
averaged over different invariant mass bins (see the legend). The
parameterization MRST01 \cite{mrst01} for the unpolarized parton
distributions is used, with $1/\beta = 0.9$ GeV$/c$. 
Curves are rescaled by a fixed $K$-factor, $K=1.6$. 
Data are from \cite{ito}.}
\label{dy400}
\end{figure}

\begin{figure} [!,h,b]
\epsfig{figure=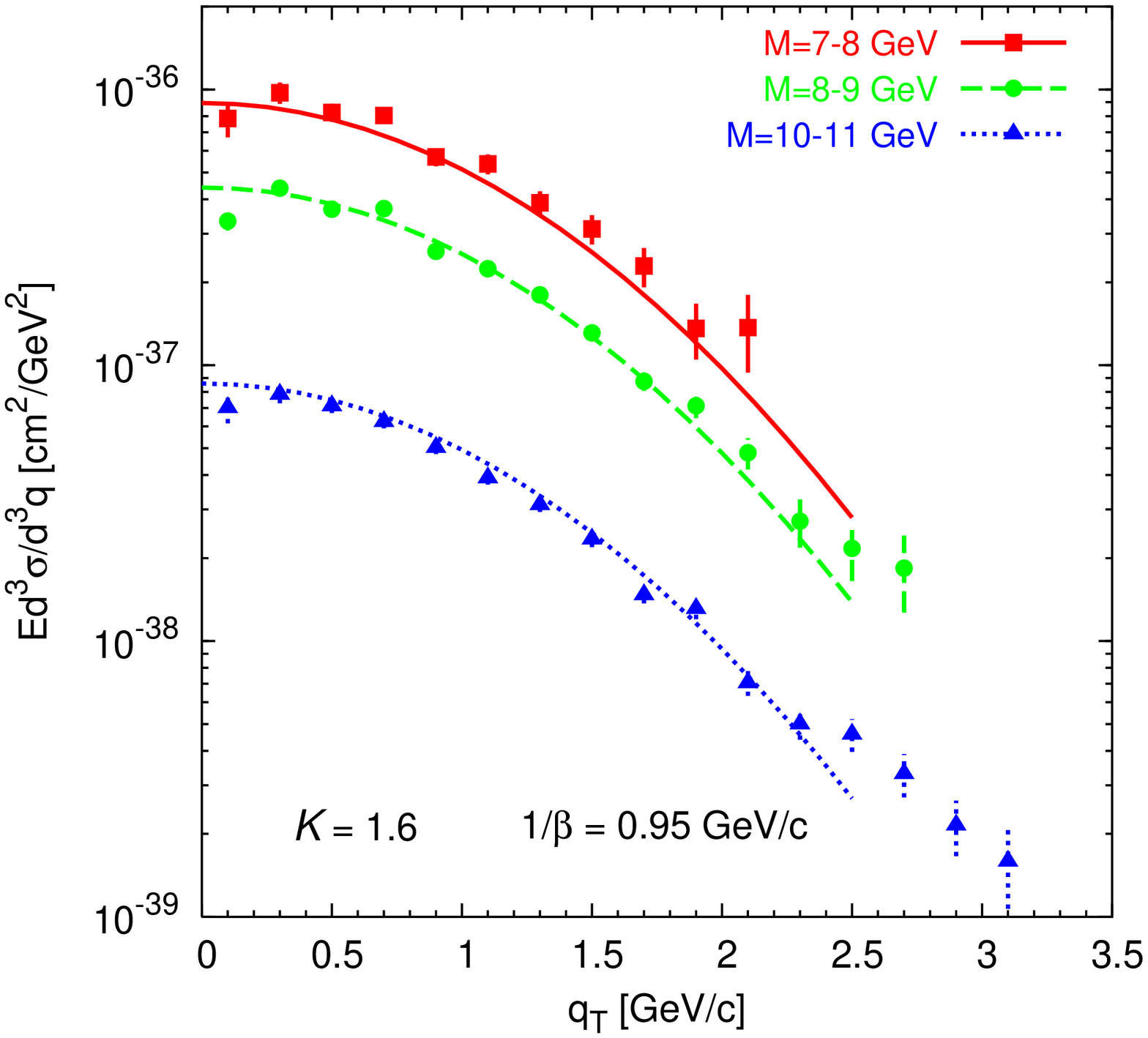,angle=-90,width=.63\textwidth}
\caption{ Invariant differential cross section for the Drell-Yan
process at $\sqrt{s}\simeq 38.8$ GeV and fixed $x_F=0.1$, as a
function of the transverse momentum of the lepton pair $q_T$ and
averaged over different invariant mass bins (see the legend). The
parameterization MRST01 \cite{mrst01} for the unpolarized parton
distributions is used, with $1/\beta = 0.95$ GeV$/c$.  
Curves are rescaled by a fixed $K$-factor, $K=1.6$. 
Data are from \cite{mor91}.}
\label{dy800}
\end{figure}

\begin{figure} [!,h,t]
\epsfig{figure=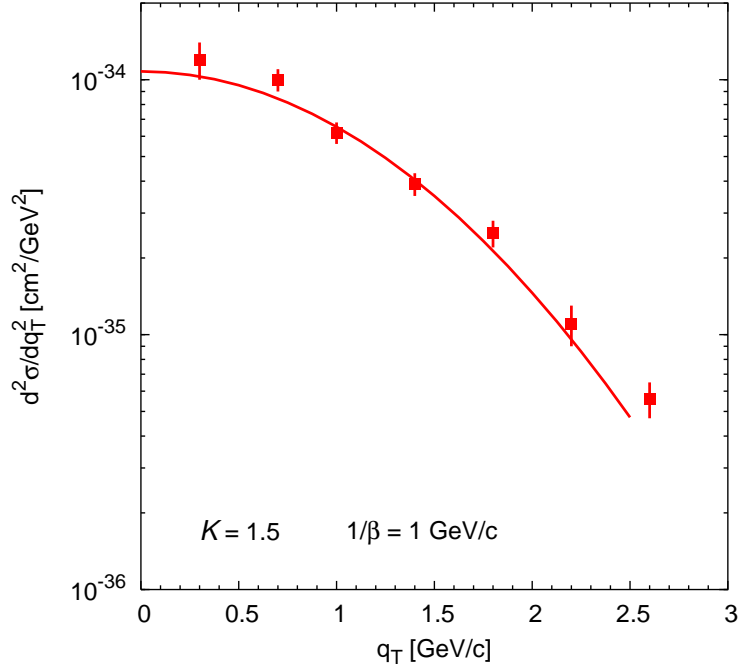,angle=-90,width=.63\textwidth}
\caption{ Differential cross section for the Drell-Yan process at
$\sqrt{s} = 62$ GeV, as a function of the transverse momentum of the
lepton pair $q_T$ and averaged over the invariant mass bin 5 GeV $< M < 8$ GeV
and over the Feynman variable bin  $-0.1<x_F<0.8$, ($x_F= 2q_L/\sqrt{s}$).  
The parameterization MRST01 \cite{mrst01}
for the unpolarized parton 
distributions is used, with $1/\beta = 1.0$ GeV$/c$.  
The theoretical curve is rescaled by a fixed $K$-factor, $K=1.5$. 
Data are from \cite{isrdy}.}
\label{dy62}
\end{figure}

In Fig.s \ref{dy200}-\ref{dy62} we compare our estimates of the Drell-Yan
invariant cross section in $p\,p$ collisions, averaged over bins of
the invariant mass $M$ and at fixed rapidity, as a function of $q_T$, 
with a collection of data from several experiments \cite{ito,mor91,isrdy}.

All curves are rescaled by proper $K$-factors varying in the range 1.5-1.8
in the different cases (see legend and caption of figures), in
agreement with NLO calculations.  The factorization scale is set to
the lepton pair invariant mass.

One can see that experimental data are described very well 
for $q_T$ values up to 2-3 GeV/$c$, where (see Fig.~\ref{dy400}) a 
power-like behaviour, entirely due to radiative effects, starts to set in.

\subsection{Direct photon production}
\label{unpg}
Let us now consider inclusive prompt photon production in $pp$ collisions
(for a compilation of data and a complete and detailed NLO analysis in
collinear pQCD of prompt photon production in hadron-hadron
interactions see \cite{vw}).

As pointed out in many papers (see for instance \cite{ow87}) intrinsic
transverse momentum effects can help in solving the discrepancy
between experimental data and LO as well as NLO collinear pQCD
calculations.  In particular, what emerges \cite{cteq-g} is that the
steep $p_T$ dependence of the measured differential
cross section cannot be explained by any
new improved PDF's (mostly for gluons); on the
contrary the introduction of intrinsic transverse momentum effects
significantly improves the comparison between theoretical calculations
and experimental data.  An attempt to give a more firm theoretical
foundation, in terms of Sudakov form factors, of the phenomenological
Gaussian $\bm{k}_\perp$ smearing has also been carried out \cite{ll}.

We employ here our LO pQCD approach (including proper $K$-factors from NLO
collinear pQCD), neglecting possible photon fragmentation
contributions, an issue related also to isolated photon cross sections
measured at extreme high-energy collider experiments (not considered
in our analysis) \cite{vw}.

Starting again from Eq.~(\ref{kqcd}), the invariant differential cross
section for the process $pp\to\gamma\,X$ then reads 
\be
\label{ppg}
E_\gamma\frac{d^3\sigma}{d^3{\bm p}_\gamma} = \sum_{ab}\! \int\!
 dx_a \, d^2\bfk_{\perp a} \, dx_b \, d^2\bfk_{\perp b} 
\hat f_{a/p}(x_a,\bfk_{\perp a}) \, \hat f_{b/p}(x_b,\bfk_{\perp b})
\, \frac{\hat s}{x_a x_b s}\,
\frac{d\hat\sigma^{ab\to \gamma d}}{d\hat t}\, \frac{\hat s}{\pi}\, 
\delta(\hat s+ \hat t + \hat u) \,, 
\ee 
where the basic partonic
processes are the Compton process $gq(\bar q)\to\gamma q(\bar q)$ and
the annihilation process $q\bar q\to\gamma g$.

By exploiting the elastic constraint, $\hat{s}+\hat{t}+\hat{u}=0$,
one of the integrations can be easily carried out, which fixes one of
the light-cone momentum fractions (e.g. $x_b$); the 5-dimensional
integral left is handled with the help of  a VEGAS Monte Carlo
routine which properly takes into account the full kinematics and all
required kinematical cuts (see also the Appendix).

The kinematical regime we are interested in is mainly the moderately
large region of $p_T$ values (1-4 GeV/$c$),
where a uniform $\bm{k}_\perp$
smearing on a steep falling $p_T$ distribution produces a
significant enhancing factor.

Similar studies (and conclusions) can be found in \cite{ww, apa98}
which however consider only photon production in the central rapidity
region. Here we extend this approach to larger $x_F$ values, in view
of its interest in studying SSA. A semi-analytical analysis with
estimates of the enhancing factor resulting from the inclusion of
intrinsic transverse momenta, in the full range of $x_F$ values, is
under completion and will be published elsewhere \cite{dmm}.

\begin{figure} [!,h,b]
\epsfig{figure=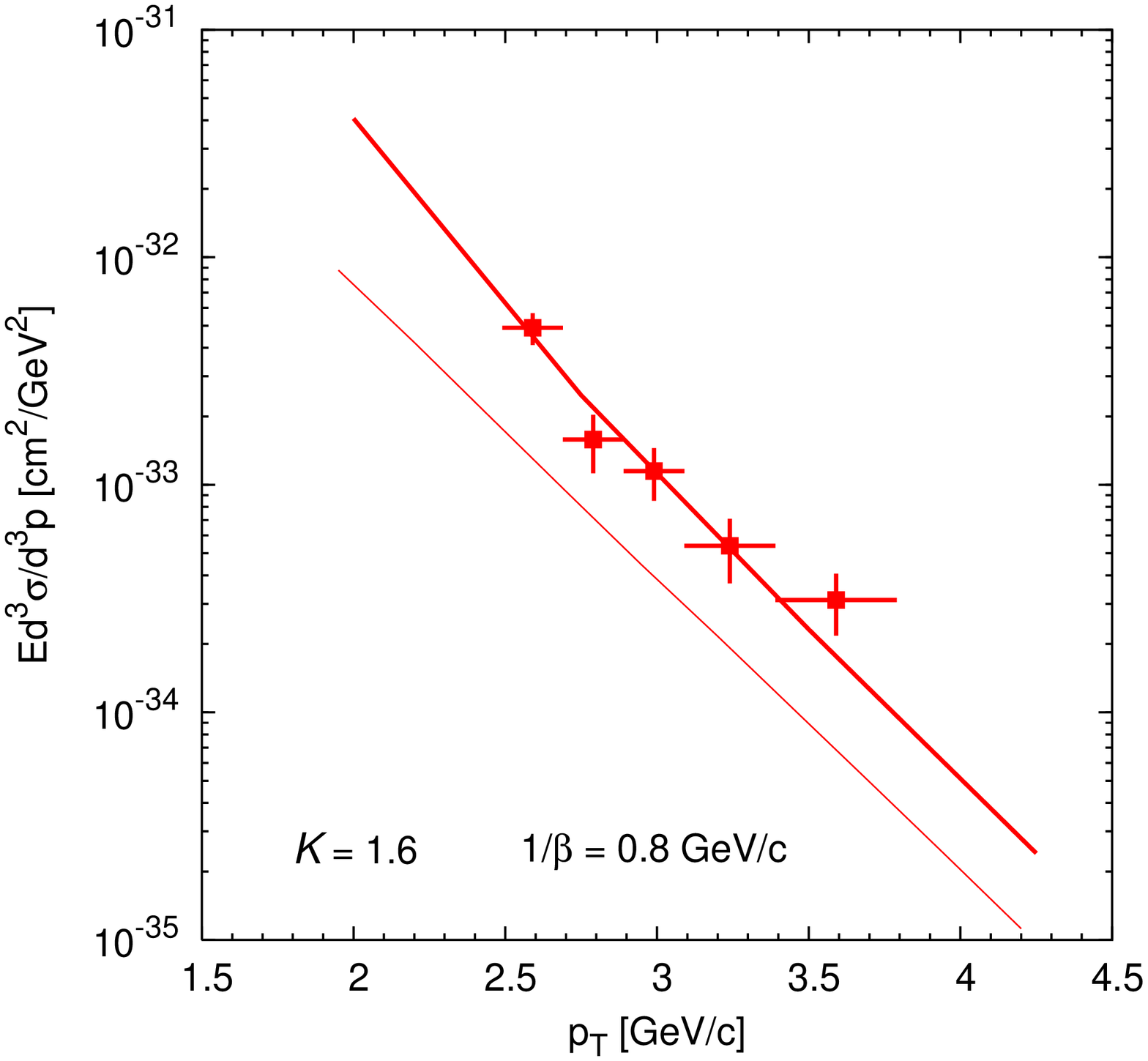,angle=-90,width=.63\textwidth}
\caption{ Invariant differential cross section for
prompt photon production in $pp$ collisions, at
$\sqrt{s} \simeq 19.4$ GeV and fixed $x_F=0$, as a function of the
photon transverse momentum $p_T$.  The parameterization MRST01
\cite{mrst01} for the unpolarized parton distributions is
used, with $1/\beta = 0.8$ GeV$/c$ (thick line). 
For comparison, the result in collinear partonic configuration
(thin line) is also shown. 
Both curves are rescaled by a fixed $K$-factor, $K=1.6$.
Data \cite{e704g} are averaged over the $x_F$ bin $-0.15<x_F<0.15$.}
\label{e704gpt}
\end{figure}

\begin{figure} [!,h,t]
\epsfig{figure=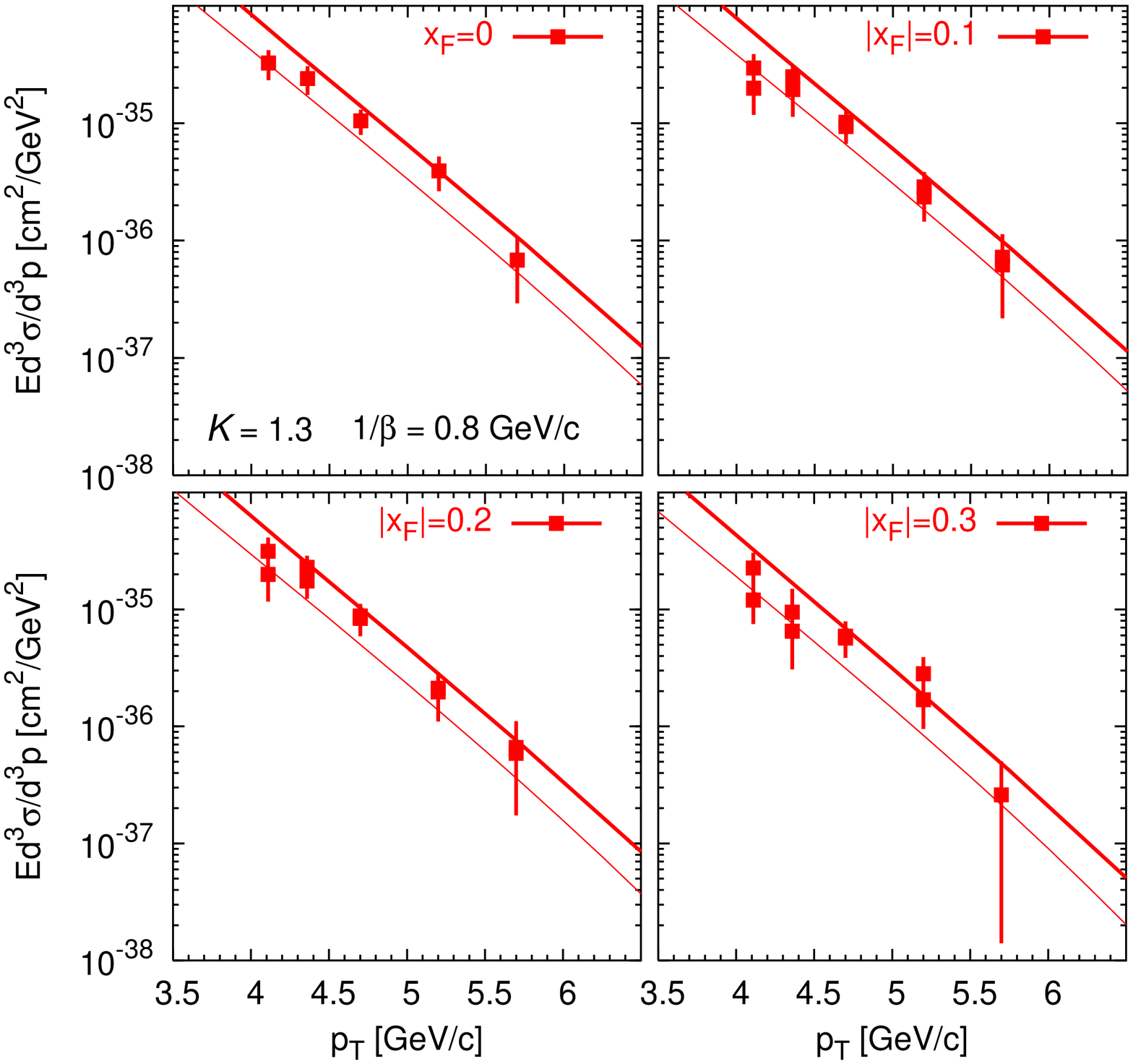,angle=-90,width=.85\textwidth}
\caption{Invariant differential cross section for
prompt photon production in $pp$ collisions, at
$\sqrt{s} \simeq 23$ GeV and different $x_F$ values, as a function of
the photon transverse momentum $p_T$.  The parameterization MRST01
\cite{mrst01} for the unpolarized parton distributions is
used, with $1/\beta = 0.8$ GeV$/c$ (thick lines). 
For comparison, the results in collinear partonic configuration
(thin lines) are also shown. 
All curves are rescaled by a fixed $K$-factor, $K=1.3$.
Data are from \cite{wa70}.}
\label{wa70pt}
\end{figure}

\begin{figure} [!,h,t]
\epsfig{figure=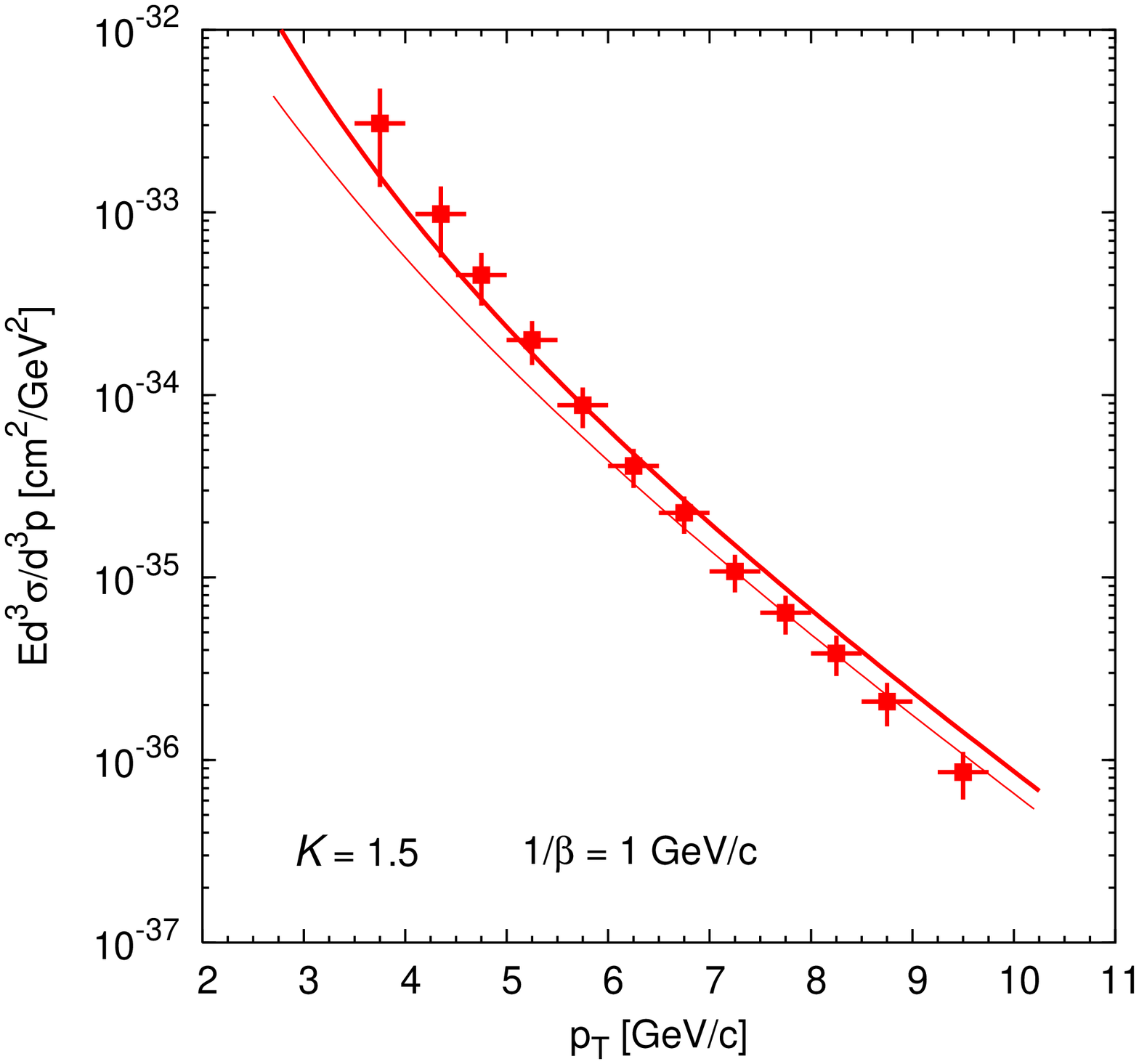,angle=-90,width=.63\textwidth}
\caption{Invariant differential cross section for
prompt photon production in $pp$ collisions, at
$\sqrt{s} = 63$ GeV and fixed rapidity $y=0$, as a function of the
photon transverse momentum $p_T$.  The parameterization MRST01
\cite{mrst01} for the unpolarized parton distributions is
used, with $1/\beta = 1.0$ GeV$/c$ (thick line). 
For comparison, the result in collinear partonic configuration
(thin line) is also shown. 
Both curves are rescaled by a fixed $K$-factor, $K=1.5$. 
Data \cite{isrg} are averaged over the rapidity bin
$-0.2<y<0.2$.}
\label{isrgpt}
\end{figure}

Our main results are compared in Fig.s~\ref{e704gpt}, \ref{wa70pt} and
\ref{isrgpt} with a representative set of experimental data for $pp$ 
collisions, both from fixed target experiments, E704 (FNAL)
\cite{e704g} and WA70 (CERN) \cite{wa70}, and from collider
experiments, R806 (ISR) \cite{isrg}. For fixed target experiments and
c.m. energies below 40
GeV we have adopted the same value of $\beta_0$ as extracted in the 
Drell-Yan process, $\beta_0 = 1.25$ (GeV/$c)^{-1}$, corresponding 
to $\langle k_\perp^2\rangle^{1/2} = 1/\beta_0 = 0.8$ GeV/$c$.
For larger energies (ISR), Fig.~\ref{isrgpt}, a
slightly smaller value of $\beta$, $\beta=1$ (GeV/$c$)$^{-1}$ seems to
better reproduce the low $p_T$ distribution. As said, we adopt the
factorization scale  $Q = p_T^*/2$, where $p_T^*$ is the photon transverse
momentum in the partonic c.m. frame. Notice that the non-collinear
parton configuration in the hadronic c.m. frame implies $p_T^* \ne
p_T$.

We use a NLO $K$-factor which decreases mildly with $p_T$ at fixed
energy ($K$= 1.6 for 1 GeV$/c<p_T<3$ GeV/$c$ and $K$ = 1.3 for 
$p_T>4$ GeV/$c$). 
This is consistent with direct calculations we performed using the
numerical code by Aurenche {\em et al.} \cite{aue} for NLO order
direct photon production.
As an example, in Fig. \ref{isrgpt}, where for the sake of simplicity
we adopted a fixed value, $K=1.5$, use of a $p_T$ dependent $K$-factor
would imply an increase of the low $p_T$ edge (where $K \simeq$ 1.9)
and at the same time a reduction of the large $p_T$ tail (where $K
\simeq$ 1.2) in our estimates, leading to a slightly better agreement
with the data than shown.

For comparison in Fig.s ~\ref{e704gpt}, \ref{wa70pt} and \ref{isrgpt}
we also show the corresponding LO, collinear pQCD results.
As expected, at very large $p_T$ the
intrinsic transverse momentum effects are negligible. In fact,
$\bm{k}_\perp$ contributions behave like $k_\perp/p_T$; moreover,
at large $p_T$ the spectrum is less steep than on the lower part,
where even a small smearing produces a big effect.

As it is clear from Fig.~\ref{wa70pt}, collinear pQCD calculations may
already give an accurate description of WA70 data ($\sqrt{s} = 23$
GeV).  On the other hand, results including $\bm{k}_\perp$ effects compare
with data equally well, while E704 data \cite{e704g},
even if at a slightly lower energy ($\sqrt{s} = 19.4$ GeV)
and at lower $p_T$, are heavily
underestimated without $\bm{k}_\perp$ effects.
 
Once more we think that the agreement between our calculations and
experimental data is good enough for our purposes.

\subsection{Inclusive pion production}
\label{unppi}

Large $p_T$ pion production in hadronic collisions is certainly
the most interesting inclusive process from many points of view.
In this case the discrepancies between collinear pQCD calculations and 
experimental data are even more significant.
This is especially true at c.m. energies below 60 GeV, where
theoretical calculations can underestimate experimental data
by a factor 10 or larger, depending also on the choice of the 
factorization scale (see also \cite{bs03}).
Again, the inclusion of intrinsic motion of partons
has been advocated in order to improve the
agreement between theory and experiment.

Most part of previous work has been addressed to inclusive pion
production in the central rapidity region \cite{pkt, apa03}.
On the other hand, as already mentioned, most of the interesting
and puzzling experimental results on SSA are in the region of
intermediate and large positive $x_F$ values; therefore a consistent
study of unpolarized cross sections in this kinematical regime and
within the same model is mandatory.

To this end, we again turn to Eq.~(\ref{kqcd}). The elastic constraint 
allows us to carry out explicitly the integration over $z$, see the 
Appendix for details. Again,  we employ a VEGAS Monte Carlo
routine to perform the remaining 8-dimensional integral with proper 
kinematical cuts.

With respect to the previous cases considered (the Drell-Yan process and
prompt photon production) a new transverse momentum dependent function
enters through the fragmentation process.  Notice also that the
unpolarized FF's are presently known with much less accuracy than the
nucleon PDF's. In particular, all available parameterizations
(with the exception of the set by Kretzer, Leader and Christova~(KLC) 
\cite{kee}) for the pion FF's are based on $e^+e^-$ data for
charged pion production, which  do not allow to
explicitly separate between $\pi^+$ and $\pi^-$ case;
this separation can only be made under further assumptions,
which remain to be tested.
The $z$ and $\bfk_\perp$ dependences in the FF are chosen according to
Eq.~(\ref{gk}).

Concerning the factorization scale, $Q$, as for the case of prompt photon
production we use $Q=\hat p_T^*/2$, where $\hat p_T^*$ is
the transverse momentum of parton $c$ in the partonic c.m. frame, that is
$\hat p_T^*= (\hat u\hat t /\hat s)^{1/2}$ (notice that in prompt
photon production the outgoing elementary particle at LO is the photon
itself).

\subsubsection{Neutral pion production}
\label{nepi}

Let us start considering inclusive neutral pion production.
Using the $\bm{k}_\perp$-dependent PDF's as
inferred from the analysis of the Drell-Yan process and comparing our
results with a large selection of cross-section data
we can extract the (flavour independent) $\beta'$ parameter entering
the pion FF, see Eq.~(\ref{gk}).
In this sub-section we will indicatively use the set by Kniehl,
Kramer, and P\"otter (KKP) \cite{kkp}.
An optimized choice of $\beta'$ seems to favour an 
explicit~$z$ dependence, which somehow spoils the simple factorized
form of Eq.~(\ref{gk}), 
\be
1/\beta'(z)=\langle\,k_{\perp \pi}^{\,2}(z)\,\rangle^{1/2}=
0.7\,z^{0.8}\,(1-z)^{0.15}\,\, {\rm GeV}/c \hspace*{2cm} {\rm [KKP]}\,.    
\label{betakkp}
\ee

Estimates of NLO corrections are taken into account by rescaling our
results by proper $K$-factors. With our choice of the
factorization scale a $K$-factor decreasing with energy and approaching
unity at very large energies \cite{wer} is to be expected.  This trend
has been also checked by direct calculations performed with the help
of the NLO numerical code for inclusive pion production by
Aversa {\em et al.} \cite{ave}.

\begin{figure} [!,h,b]
\epsfig{figure=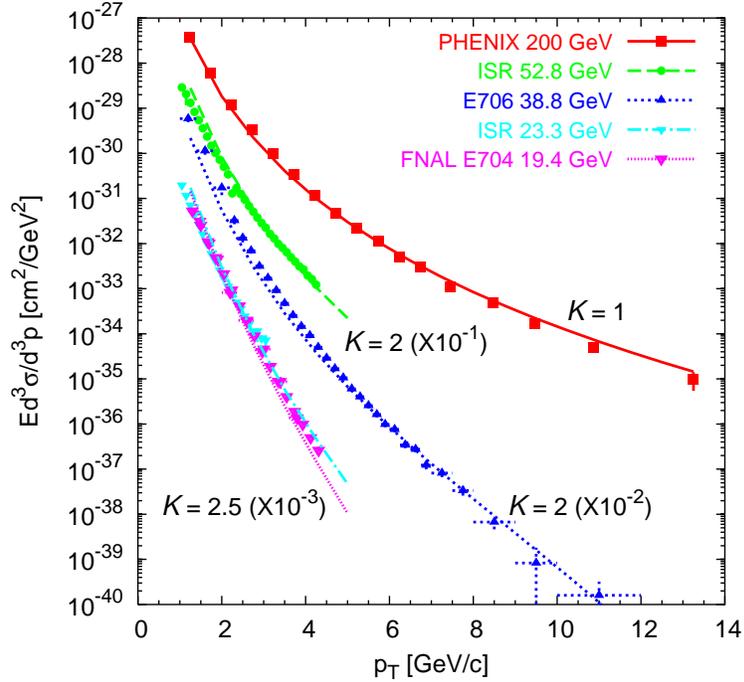,angle=-90,width=.63\textwidth}
\caption{Invariant differential cross section for 
inclusive $\pi^0$ production in $pp$ collisions, at
different c.m.~energies (see the legend) and fixed rapidity $y=0$, as a
function of the pion transverse momentum $p_T$.   
We use the parameterization MRST01 \cite{mrst01} for the unpolarized 
parton distributions, with $1/\beta = 0.8$ GeV$/c$, and the
parameterization KKP \cite{kkp} for the unpolarized fragmentation
functions, with $\beta'$ given in Eq.~(\ref{betakkp}).   
$K$-factors for each case are shown (some data sets and their
corresponding curves are further rescaled for clarity).  Data are from
\cite{ada96, dona, apa03, isrpi, adl03}.}
\label{pi0y0}
\end{figure}

In Fig.~\ref{pi0y0} we present our estimates of the invariant
differential cross section for the process $p\,p\to\pi^0\,X$ at
different energies vs. $p_{T}$ and at central rapidity, $y=0$. After
rescaling each curve by the proper $K$-factor, ranging from 2.5 to 1,
we get reasonable agreement with experimental data 
\cite{dona,ada96,isrpi,adl03} over a large range of energies (from 20
to 200 GeV).
 
\begin{figure} [!,h,t]
\epsfig{figure=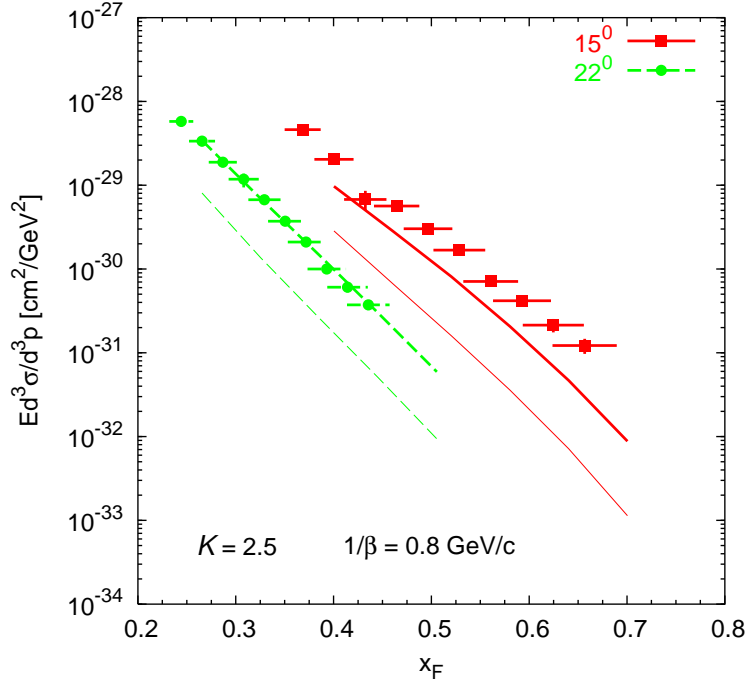,angle=-90,width=.63\textwidth}
\caption{Invariant differential cross section 
for inclusive $\pi^0$ production in $pp$ collisions, at
$\sqrt s =23.3$ GeV and for two c.m.~scattering angles, as a function of
$x_F$.  
We use the parameterization MRST01 \cite{mrst01} for the unpolarized 
parton distributions, with $1/\beta = 0.8$ GeV$/c$, and the
parameterization KKP \cite{kkp} for the unpolarized fragmentation
functions, with $\beta'$ given in Eq.~(\ref{betakkp}) (thick lines).   
For comparison, the results in collinear partonic configuration (thin
lines) are also shown.
All curves are rescaled by a fixed $K$-factor, $K=2.5$. 
Data are from \cite{isrpi}.}
\label{isrpi1}
\end{figure}

\begin{figure} [!,h,b]
\epsfig{figure=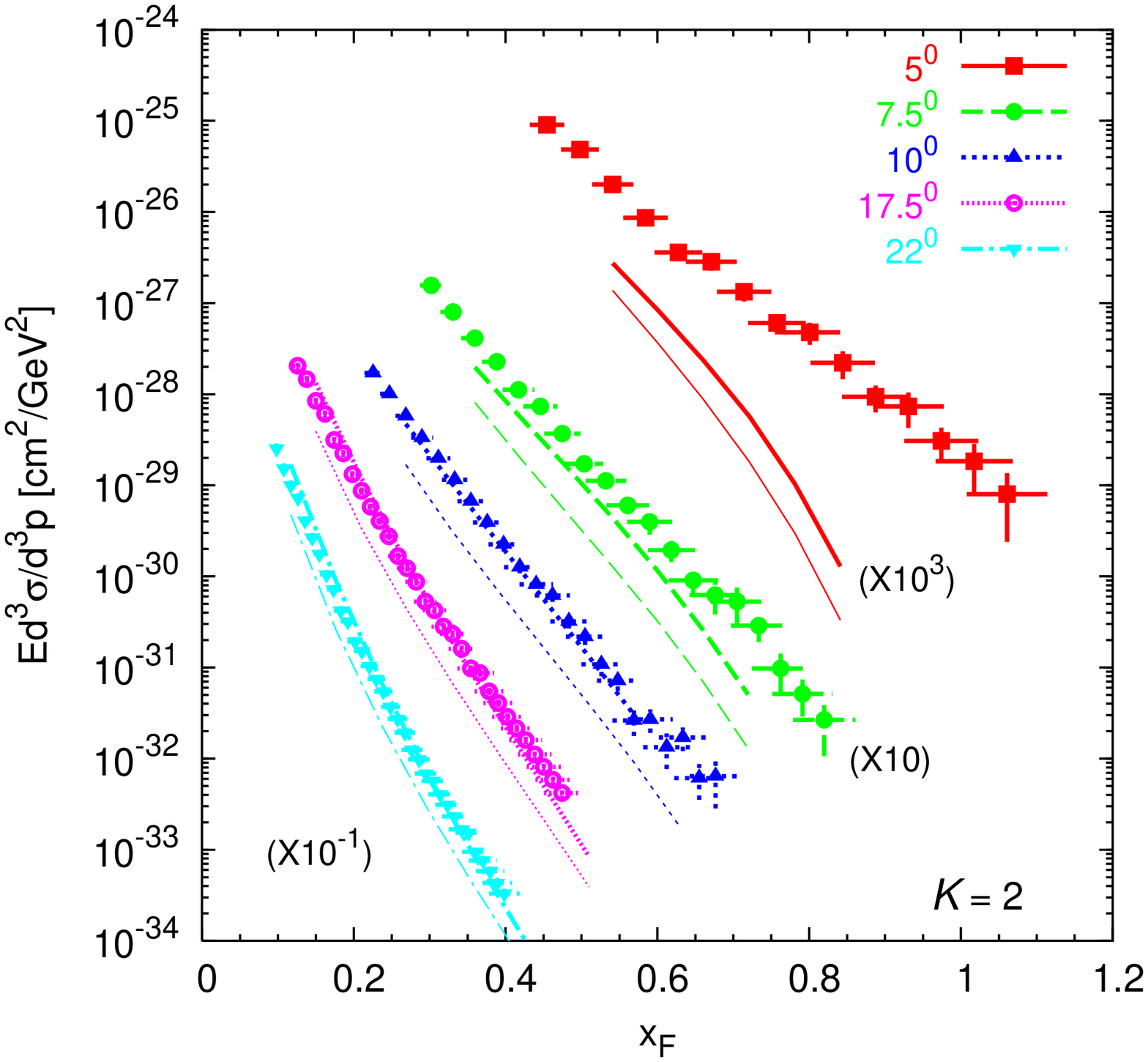,angle=-90,width=.63\textwidth}
\caption{Invariant differential cross section 
for inclusive $\pi^0$ production in $pp$ collisions, at
$\sqrt s = 52.8$ GeV and for different c.m.~scattering
angles (see the legend), as a function of $x_F$.  
We use the parameterization MRST01 \cite{mrst01} for the unpolarized 
parton distributions, with $1/\beta = 0.8$ GeV$/c$, and the
parameterization KKP \cite{kkp} for the unpolarized fragmentation
functions, with $\beta'$ given in Eq.~(\ref{betakkp}) (thick lines).   
For comparison, the results in collinear partonic configuration (thin
lines) are also shown.
All curves are rescaled by a fixed $K$-factor, $K=2$ (some data sets with their
corresponding curves are further rescaled for clarity). 
Data are from \cite{isrpi}.}
\label{isrpi2}
\end{figure}

A bit more problematic is the case of neutral pion production in the very
forward region (very small scattering angles), as shown in Figs.~\ref
{isrpi1} and \ref{isrpi2}.  In particular for $\theta< 10^0-15^0$ even
with the inclusion of intrinsic $\bm{k}_\perp$ effects the discrepancy
between theory and data at ISR energies \cite{isrpi}, $\sqrt{s}$ =
23.3 and 52.8 GeV, is still large. Nonetheless in these plots the LO collinear
calculations (rescaled by the same $K$-factors) show how
important the enhancing factor due to $\bm{k}_\perp$ contributions can be.

\begin{figure} [!,h,t]
\epsfig{figure=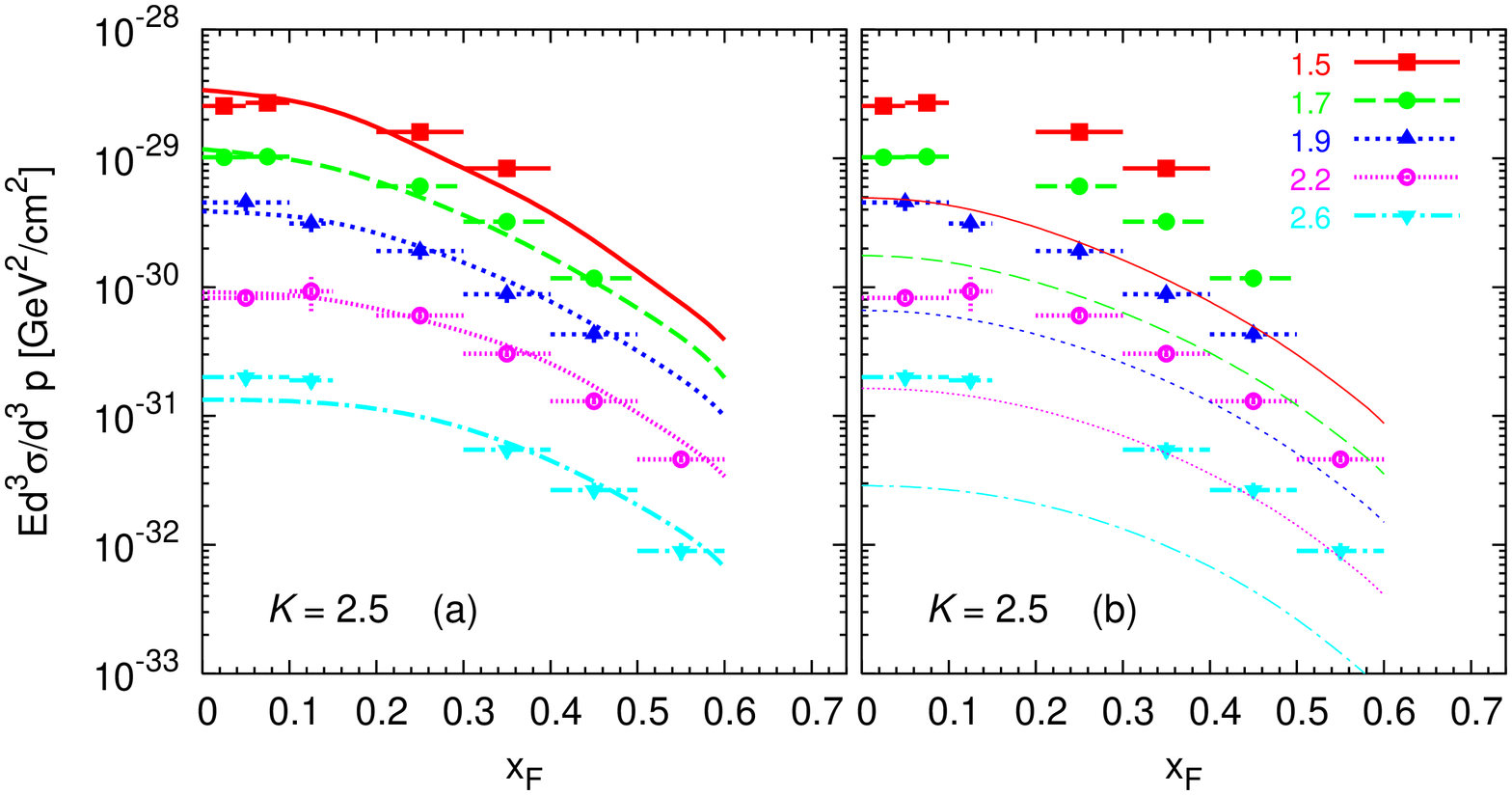,angle=-90,width=.9\textwidth}
\caption{Invariant differential cross section for 
inclusive $\pi^0$ production in $pp$ collisions, at
$\sqrt s = 19.4$ GeV and for different $p_T$ values (in GeV$/c$, see the
legend), as a function of $x_F$.  
We use the parameterization MRST01 \cite{mrst01} for the unpolarized 
parton distributions, with $1/\beta = 0.8$ GeV$/c$, and the
parameterization KKP \cite{kkp} for the unpolarized fragmentation
functions, with $\beta'$ given in Eq.~(\ref{betakkp}) 
(plot (a), thick lines).   
For comparison, the corresponding results in collinear partonic configuration 
(plot (b), thin lines) are also shown.
All curves are rescaled by a fixed $K$-factor, $K=2.5$.  
Data are from \cite{dona}.}
\label{fnal}
\end{figure}

In Fig.~\ref{fnal} we compare our estimates, rescaled by a $K$-factor
$K=$ 2.5, with data at $\sqrt{s} = 19.4$ GeV \cite{dona}; here 
fair agreement can be reached over a broad region of $x_F$ 
away from 0, and at different $p_T$ values.
To our knowledge, these data are the only ones available covering
almost the same kinematical region of the well-known E704 results on 
pion SSA \cite{e704}.

\begin{figure} [!,h,b]
\epsfig{figure=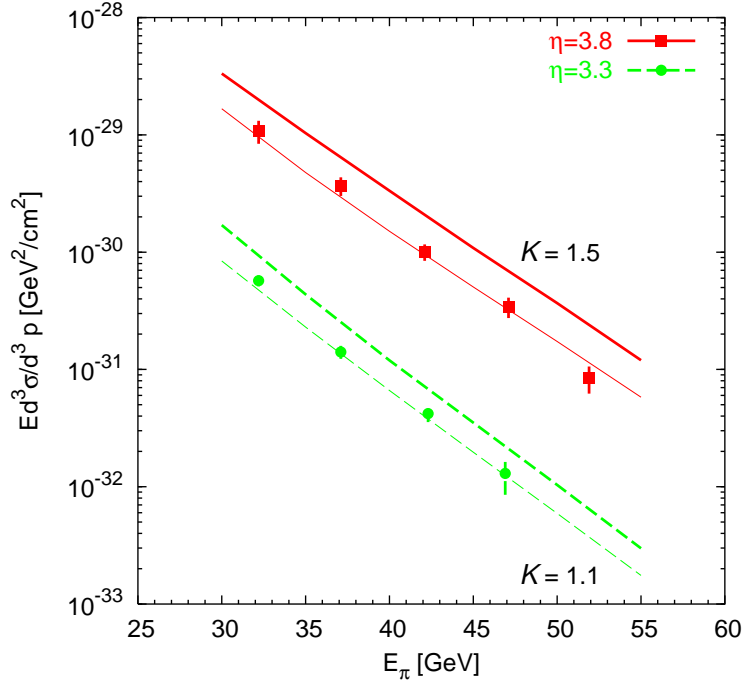,angle=-90,width=.63\textwidth}
\caption{Invariant differential cross section 
for inclusive $\pi^0$ production in $pp$ collisions, at
$\sqrt s = 200$ GeV and for two values of the pseudo-rapidity, $\eta$, 
as a function of the pion energy $E_{\pi}$.  
We use the parameterization MRST01 \cite{mrst01} for the unpolarized 
parton distributions, with $1/\beta = 0.8$ GeV$/c$, and the
parameterization KKP \cite{kkp} for the unpolarized fragmentation
functions, with $\beta'$ given in Eq.~(\ref{betakkp}) (thick lines).   
For comparison, the results in collinear partonic configuration (thin
lines) are also shown. $K$-factors for the two cases are shown. 
Data are from \cite{ada03}.}
\label{star}
\end{figure}

Recent data from STAR \cite{ada03} in comparable $x_F$ and $p_T$
ranges but at much larger energies are shown in Fig.~\ref{star}, together with
our theoretical curves, rescaled by proper $K$-factors.
Let us remind that we use for each curve a fixed $K$-factor, whereas
use of a $p_T$-dependent $K$-factor would imply an increase of the
low $E_\pi$ edge and a reduction of the large $E_\pi$ tail.
In this case we overestimate the data,
but still a reasonable description is obtained.  
Let us notice that, since data on pion SSA come mainly from E704
experiment, in performing our optimized choice for $\beta'$, we have given a
slight preference to unpolarized cross-section data in the same
kinematical region \cite{dona}. A smaller value of $1/\beta'$  would
lead to a better agreement with STAR data without spoiling the good
description of FNAL experimental results \cite{dona}. 
Let us also recall, in this
respect, that the results presented cover a huge spectrum of different
kinematical situations at different c.m. energies.

We have also checked that the pion FF set by Kretzer \cite{kre} with a
corresponding proper choice of the $\beta'$ parameter,
\be
1/\beta'(z)=\langle\,k_{\perp \pi}^{\,2}(z)\,\rangle^{1/2}=
1.4\,z^{1.2}\,(1-z)^{0.35}\,\, {\rm GeV}/c \hspace*{2cm} {\rm [Kretzer]}\,,  
\label{betakre}
\ee
gives similar results.
For the processes under consideration here the KLC set
\cite{kee} is practically equivalent to the set of Ref. \cite{kre}.

\subsubsection{Charged pion production}
\label{chpion}

The study of inclusive charged pion production needs some additional
caution; as it was already mentioned  the available pion FF sets come
from fits to experimental data on hadron production in $e^+e^-$
annihilation, where only the sum of charged pions is detected.
In order to get flavour (or charge) separation one therefore has to
rely on some extra assumptions: for instance Kretzer \cite{kre}
imposes, besides isospin symmetry, the following behaviour of the
non-leading quark FF's at the input scale ($\mu_0^2$): 
\bea
\label{kresep}
D_{\pi^+/d}(z,\mu_0^2) & = & (1-z)\, D_{\pi^+/u}(z,\mu_0^2) \,,\\ 
D_{\pi^+/s}(z,\mu_0^2) & = & D_{\pi^+/d}(z,\mu_0^2) \,.  
\eea
Under proper pQCD evolution, a FF set for $\pi^0$ and $\pi^\pm$ can
then be extracted.  In other parameterizations, like the KKP set
\cite{kkp}, no additional flavour separation assumption is made and
only a FF set for neutral pions is available.

Clearly the relative weight between valence (or leading)
and sea (or non-leading) quark FF's can play a relevant role
in determining the relative yields of $\pi^+$ and $\pi^-$.
 
For these reasons, we model two charged pion FF sets, starting from
the neutral pion KKP parameterization.  In one case (KKP-1) we assume
that the non-leading (sea) quark FF's into a charged pion are all
equal to the strange quark FF into a neutral pion: this leads to a
strongly valence-dominated fragmentation mechanism at large $z$
values.  A second case (KKP-2) is defined along the lines of Kretzer
set. 
More precisely, assuming isospin symmetry, $D_{\pi^-/u} = D_{\pi^+/d}$
and charge conjugation invariance, $D_{\pi^-/u} = D_{\pi^+/\bar u}$, we
define for KKP-1 
\beq
\label{KKP1}
D_{\pi^+/d}(z) = D_{\pi^0/s}(z)\>,
\eeq
that implies
\beq
D_{\pi^+/u}(z) = 2
D_{\pi^0/u}(z) - D_{\pi^0/s}(z) \hspace*{4cm} ({\rm KKP-}1) \,.
\eeq 
For KKP-2 set instead we impose 
\beq
\label{KKP2}
D_{\pi^+/d}(z) = (1-z)^a D_{\pi^+/u}(z)\>,
\eeq 
that implies 
\beq
D_{\pi^+/u}(z) = 2 D_{\pi^0/u}(z) / [1+(1-z)^a]
\hspace*{3.5cm} ({\rm KKP-}2) \,,
\eeq 
where the parameter $a$, chosen
reasonably around 1-1.5, will be fixed to $a=1.3$ in the
following. Clearly, these sets leave unchanged,  and equal to those of
the KKP set, the FF's for neutral pions.

Let us remark that the analysis of Ref. \cite{kee}, based on a
simultaneous fit to $e^+e^-$ and semi-inclusive DIS results which in
principle allows flavour separation, indicates a relatively small
suppression of non-leading vs. leading quark FF's, as already assumed in
Kretzer set.
 
\begin{figure} [!,h,t]
\epsfig{figure=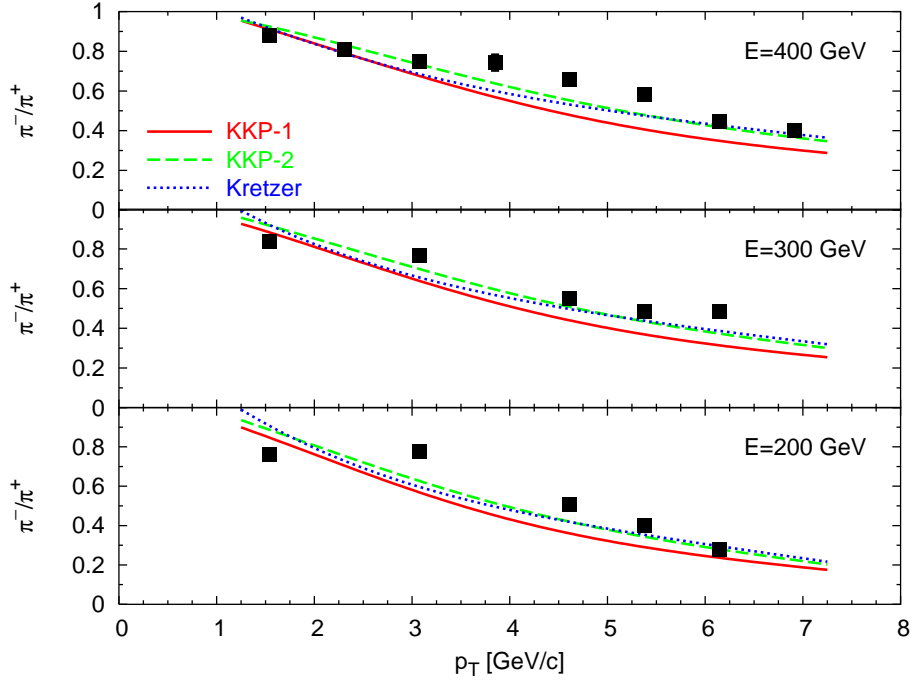,angle=-90,width=.65\textwidth}
\caption{ Ratios of invariant differential cross sections
for inclusive charged pion production in $pp$ collisions,
at different LAB energies and fixed scattering angle
$\theta = 77$ (mrad), as a function of $p_T$.   
We use the parameterization MRST01 \cite{mrst01} for the unpolarized 
parton distributions, with $1/\beta = 0.8$ GeV$/c$, 
 and three different sets for charged pion FF's with the
corresponding $\beta'$ parameters as in Eq.~(\ref{betakkp}) and
(\ref{betakre}) (see text). Data are from \cite{FNALpm}.}
\label{ratiomp}
\end{figure}

In Fig.~\ref{ratiomp} we show our estimates for the ratio of $\pi^-$
and $\pi^+$ yields and compare them with FNAL experimental data
\cite{FNALpm} for three different energies and at
scattering angle $\theta= 77$ mrad (at these energies, this
corresponds almost to $x_F=0$) as a function of $p_T$.  Results using
the three different options for pion FF's discussed above (Kretzer,
KKP-1 and KKP-2) are shown.  Since all pion FF sets adopted assume
that the FF for a neutral pion coincides with the average of those for
charged pions, our estimates for $\pi^0$ cross sections, Fig.s 
\ref{pi0y0}-\ref{star},
together with the results of Fig.~\ref{ratiomp}, give complete
information for the charged pion sector. Therefore, we do not
explicitly show our results for charged pion cross sections,
which are in good agreement with data.

{}From the ratio of $\pi^-$, $\pi^+$ yields we see that both Kretzer and
KKP-2 sets give a reasonably good description of the experimental
data, while KKP-1 set seems to underestimate the data, in
particular at large $p_T$. Let us however remark that the KKP-1 and
KKP-2 sets are model sets based on some assumptions which remain to be tested.
We propose them here as representative of two very different
behaviours with respect to the relative weight of leading and
non-leading pion FF's. As such, our results are not intended to give
preference to any set of FF's over the others.

\section{Transverse single spin asymmetries and Sivers effect}
\label{tssa}

We come now to the other main subject of this paper.  Parton
intrinsic transverse momentum effects, as we have seen in the previous
sections, can play a crucial role in reducing the discrepancies
between pQCD estimates and experimental data for unpolarized cross
sections in inclusive hadron production.  What is more important from
our point of view, they can be an essential ingredient in
understanding some spin phenomena in the framework of pQCD.
The large SSA observed in many reactions like 
$A^{\uparrow}\,B \to C\,X$ or $A\,B \to H^{\uparrow}\,X$ 
(where $H$ is typically a hyperon) can be
described by certain spin-dependent effects generated by soft
mechanisms in the presence of intrinsic transverse momentum.  These
effects and several applications to different processes, including
various possible mechanisms of this sort, have been discussed in a
series of papers \cite{noiS,noiC}, keeping only dominant
$\bm{k}_\perp$ contributions and under some simplifying assumptions
about parton kinematics and $\bm{k}_\perp$ distributions.

Let us then consider the transverse single spin asymmetry 
\beq
\label{an}
A_N = \frac{d\sigma^\uparrow - d\sigma^\downarrow} 
{d\sigma^\uparrow + d\sigma^\downarrow} 
\eeq 
where $d\sigma^{\uparrow,\downarrow}$ stands for the
invariant differential cross section for the process 
$A^{\uparrow,\downarrow} + B \to C + X$,
and $\uparrow,\downarrow$ indicates the transverse direction with
respect to the production plane (corresponding to the $\pm Y$
direction in a reference system where the initial polarized hadron
moves along the $+Z$ direction and the observed hadron lies in the
$+X$-$Z$ half-plane).
As it was already mentioned, at leading twist  
there are three possible soft mechanisms (and
three corresponding new spin and $\bm{k}_\perp$ dependent PDF's/FF's)
contributing to this process, see Eq.s~(\ref{delf2}), (\ref{deld2}),
often referred to as ``odd under naive time reversal'': \\ 
a) The Sivers effect (and distribution function) \cite{siv},
corresponding to the possible azimuthal dependence (around the
light-cone direction of the parent nucleon) of the number density of 
unpolarized quarks (or gluons) inside a transversely polarized nucleon. \\ 
b) The Collins effect (and fragmentation function) \cite{col},
corresponding to the azimuthal dependence (around the light-cone
direction of the fragmenting parton) of the number density of
unpolarized hadrons resulting from the fragmentation of a transversely 
polarized quark (or linearly polarized gluon). \\ 
c) The Boer-Mulders effect (and distribution function) \cite{muld,boer}, 
corresponding to the azimuthal dependence (around the light-cone
direction of the parent nucleon) of the number density of transversely 
polarized quarks (linearly polarized gluons) inside an unpolarized nucleon.

When considering the production of a transversely polarized, spin-1/2
hadron (like e.g. in the process $A\,B\to H^{\uparrow}\,X$)
another leading twist, $T$-odd effect
has to been taken into account in the fragmentation process:\\
d) The so-called polarizing fragmentation function \cite{abdm1}, 
corresponding to the azimuthal dependence (around the light-cone 
direction of the fragmenting parton) of the number density of 
transversely polarized  hadrons resulting from the fragmentation 
of an unpolarized quark (or gluon).

It should be noted that in the pure parton model, where
partons are treated as physical, massless free particles, all these
effects, when described as totally independent, factorized processes, 
vanish \cite{ell}.

Let us remind that these functions may be defined also for gluons;
in this case, however, instead of transverse quark polarization we
must refer to linear gluon polarization \cite{rodr,abdlmm}.

All of the above spin effects a)--c) could contribute in principle to
the process $p^\uparrow p\to \pi+X$, but, at the present stage, a
complete phenomenological analysis is out of reach and perhaps not very
significant, due to the relatively scarce experimental information
available. On the other hand, it is very important to verify whether
any of these possible contributions, when taken alone, may contribute
and to what extent to the SSA observed, extracting even gross features
of the corresponding PDF's/FF's.

In the following we will then concentrate on SSA in inclusive pion and 
photon production generated by the Sivers effect \cite{siv}.
SSA in Drell-Yan processes within the same approach have been considered
in Ref.~\cite{adm1}. For a complementary phenomenological study along
the same lines, but invoking only Collins effect, see Ref.~\cite{abdlm}.

A more formal and complete analysis including all possible 
leading-order, leading-twist contributions for single and double spin
asymmetries in inclusive pion production is currently under
way and will be presented elsewhere \cite{abdlmm}.

Let us then start from the most general expression for the number density
of unpolarized quarks $q$ (or analogously gluons), with light-cone
momentum fraction $x$ and
transverse momentum $\bfk_{\perp}$, inside a proton with transverse
polarization $\bfP$ and three-momentum $\bfp$. One can write
\be 
\hat f_{q/\pup}(x, \bfk_{\perp}) = \hat f_{q/p}(x, k_{\perp}) +
\frac{1}{2} \, \Delta^N f_{q/\pup}(x, k_{\perp}) \> \hat{\bfP} \cdot
(\hat{\bfp} \times \hat{\bfk}_{\perp})\>, 
\label{polden} 
\ee
which implies 
\bea 
\hat f_{q/p}(x, k_{\perp}) &=& \frac{1}{2}[ \hat f_{q/\pup}(x,
\bfk_{\perp}) + \hat f_{q/\pdown}(x, \bfk_{\perp})]
\label{f+f} \,,\\
\Delta^Nf_{q/\pup}(x, \bfk_{\perp}) &=& \hat f_{q/\pup}(x,
\bfk_{\perp})-\hat f_{q/\pdown}(x, \bfk_{\perp})
= \Delta^Nf_{q/\pup}(x,k_{\perp})\,\sin(\phi_{P}-\phi_{k_\perp})\,,
\label{f-f}
\eea 
where $\phi_{P}$ and $\phi_{k_\perp}$ are respectively the azimuthal
angle of the proton (transverse) polarization vector and of
$\bm{k}_\perp$. Eq.~(\ref{f+f})
gives the unintegrated, $k_\perp$-dependent unpolarized PDF, while 
Eq.~(\ref{f-f}) defines the Sivers PDF, $\Delta^Nf_{q/\pup}(x,k_{\perp})$.
Another common notation for the Sivers function is 
$f_{1T}^{\perp}(x,k_\perp)$ \cite{muld}; it should be noted that the 
two definitions are not completely equivalent, the exact relation being
$\Delta^Nf_{q/\pup}(x,k_{\perp})=
(-2k_\perp/M_p)\,f_{1T}^{\perp}(x,k_\perp)$.

In order to perform numerical estimates and carry out a comparison
with available experimental data we introduce a simple model for the
Sivers function (see also Ref. \cite{adm1} for more details), similar
to what was previously done for the so-called polarizing fragmentation
function in Ref. \cite{abdm1}.

Analogously to the case of the unpolarized PDF's and FF's,
Eq.~(\ref{gk}), we consider a simple factorized form for the Sivers function:
\be 
\Delta^Nf_{q/\pup}(x, k_\perp) = \Delta^Nf_{q/\pup}(x) \,
h(k_\perp)\>.
\label{modd}
\ee

In order to satisfy the positivity bound (see Eq.s~(\ref{f+f}), (\ref{f-f}))
\be 
|\Delta^N f_{q/\pup}(x, k_{\perp})| \leq 2 \,\hat f_{q/p}(x, k_{\perp})\,, 
\qquad \forall\, x,\,k_{\perp} \,,
\label{posb}
\ee 
we put
\bea 
\Delta^Nf_{q/\pup}(x) & = & 2\,{\mathcal N}_q(x)\,f_{q/p}(x)\,
\label{dnf} \\
h(k_\perp) & = &{\mathcal H}(k_\perp) \,\frac{\beta^2}{\pi}\,
e^{-\beta^2 \, k_\perp^2} \>,
\label{hk}
\eea
and we simply need to require
\be 
|{\mathcal N}_q(x) \, {\mathcal H}(k_\perp)| \le 1 \quad\quad
\forall\, x,\,k_\perp \,. 
\label{bound2} 
\ee
We actually impose ${\mathcal N}$ and ${\mathcal H}$ to be separately
smaller than unity in magnitude, by choosing simple functional forms
and dividing each of them by its maximum value:
\be 
{\mathcal N}_q(x) = N_q\,x^{a_q}(1-x)^{b_q}\,
\frac{(a_q+b_q)^{(a_q+b_q)}}{a_q^{a_q}\,b_q^{b_q}}\,,\quad 
|N_q|\leq 1\,
\label{nqx}
\ee
\be 
{\mathcal H}(k_\perp)=\sqrt{2\,e\,(\alpha^2-\beta^2)}\,
k_\perp\,\exp\,\left[\,-(\alpha^2-\beta^2)\,k_\perp^2\,\right]\, ,
\quad \alpha > \beta\>.
\label{hhk}
\ee

Notice that ${\mathcal H}(k_\perp)$ must vanish at least like
$k_\perp$ for $k_\perp\to 0$. Eq.s (\ref{hk}) and (\ref{hhk}) imply:
\be 
\Delta^Nf_{q/\pup}(x, k_\perp) = 2 \, {\mathcal N}_q(x) \,
f_{q/p}(x) \, \frac{\beta^2}{\pi} \, \sqrt{2\,e\,(\alpha^2 -
\beta^2)}\, k_\perp \, e^{-\alpha^2 k_{\perp}^2} \>. 
\label{del2} 
\ee
Since the positivity bound requires $\alpha > \beta$, one can write 
$\alpha^2=\beta^2/r$, where $r$ is a positive parameter smaller than
one. We will keep this last parameter flavour and $x$-independent.

Let us stress again that in principle one can also define a
gluon Sivers function, for which the same expressions would be valid.
In order to reduce the number of free parameters and assuming in first
approximation that transverse spin-$\bfk_\perp$ effects are
valence-dominated (certainly a reasonable assumption for particles
produced at large positive values of $x_F$), we will restrict here to 
consider valence quarks, neglecting (for what concerns the Sivers
effect only) possible contributions from sea quarks and gluons.
We have explicitly checked that, for the processes under study,
even maximizing the corresponding Sivers functions
these contributions could play a  role
only at low $x_F$, where $A_N$ is almost negligible
(see also Ref.~\cite{noiD}).

Under these assumptions we are left with seven free parameters which
completely fix the Sivers function as given in Eq.~(\ref{del2}); six
for ${\mathcal N}_q(x)$ in Eq.~(\ref{nqx}), with $q=u,d$, plus
$r=\beta^2/\alpha^2$ in Eq.~(\ref{hhk}).

\subsection{ Sivers effect in $p^\uparrow p\to\pi X$}
\label{ssapi}

By comparing our calculations for SSA in pion production with available
experimental data we now try to fix the set of parameters entering the
Sivers function.
  
According to Eq.~(\ref{kqcd}), the numerator of the SSA,
Eq.~(\ref{an}), in terms of Sivers effect alone reads
\bea 
E_\pi\frac{d\sigma^\uparrow}{d^3\bfp_\pi} -
E_\pi\frac{d\sigma^\downarrow}{d^3\bfp_\pi} &=& \sum_{a,b,c,d} \int
dx_a\,d^2\bfk_{\perp a}\,dx_b\,d^2\bfk_{\perp b}\,dz \,d^3\bfk_\pi \,
\delta(\bfk_\pi\cdot \hat{\bfp}_c)\, 
\Delta^N f_{a/\pup}(x_a, \bfk_{\perp a}; Q^2) \> \hat
f_{b/p}(x_b, \bfk_{\perp b}; Q^2) \nonumber \\ 
&\times& \> 
\frac{\hat s}{x_a x_b s}\> 
\frac{d\hat\sigma^{ab \to cd}}{d\hat t}(x_a,x_b,
\hat s, \hat t, \hat u) \> 
\frac{\hat s}{\pi}\,\delta(\hat s + \hat t + \hat u) \>
\frac{1}{z^2}\,J(z,|\bfk_\pi|) \,
\hat D_{\pi/c}(z, \bfk_\pi; Q^2) \>.
\label{sivgen}
\eea
The denominator of the SSA, $d\sigma^\uparrow+d\sigma^\downarrow$,
is simply given by twice the unpolarized cross section, Eq.~(\ref{kqcd}).

Let us remind that we consider the kinematics of the reaction in the 
c.m.~of the colliding protons, where $\uparrow,\downarrow$ indicates the 
transverse direction with respect to the production plane
(corresponding to the $\pm Y$ direction when the initial polarized
proton moves along the $+Z$ direction and the observed pion lies in
the $+X$-$Z$ half-plane).
According to Eq.~(\ref{f-f}), in this frame $\phi_P=\pi/2$, and
\beq 
\Delta^{\!N}f_{a/\pup}(x_a, \bfk_{\perp a}) =
\Delta^{\!N}f_{a/\pup}(x_a, k_{\perp a})\, \cos(\phi_{k_{\perp a}})\,. 
\eeq 

As already discussed in section \ref{unppi},
energy-momentum conservation in Eq.~(\ref{sivgen}) allows us to carry out
e.g. the integration over $z$ (see also the Appendix) leaving us with an
8-dimensional integral.  This makes unpractical and very
time-consuming a complete best-fit procedure over all seven free
parameters. Moreover, a complete analysis of this kind would probably
be premature and not very significant at this stage. Therefore, we
here adopt a less rigorous approach and try to reproduce the
experimental data, extracting an optimized
choice for the Sivers function parameters from a limited set of
choices in the available parameter phase-space. \\ 
To this end, we
consider the experimental data for pion SSA of the E704
Collaboration \cite{e704}, at $\sqrt{s}=19.4$ GeV.  These data cover a
large range of (positive) values for $x_F$, with $p_T$ in the range
[0.7--2.0] GeV/$c$. In what follows, we will use for our calculations
an averaged value of $p_T=1.5$ GeV/$c$. Although this value might be
slightly higher than the effective averaged $p_T$ of the data set, it
is well indicative of the kinematic regime analysed.

We start by noticing what the E704 data show: at large positive $x_F$
the SSA, $A_N$, for charged pions is large and almost of the same
magnitude, but opposite in sign for $\pi^+$ and $\pi^-$
(positive for $\pi^+$). In order to
describe these results in our approach invoking the Sivers effect alone we
need a positive (negative) $up$ ($down$) Sivers function,
$\Delta^{\!N}f_{q/p^\uparrow}(x,k_\perp)$.
This means that the non-leading flavour 
contributions (e.g. $d\to\pi^+)$) enter with opposite sign (all other terms in
Eq.~(\ref{sivgen}) giving positive contributions) with respect to the
leading ones (e.g. $u\to\pi^+$) in each charged pion SSA asymmetry. 
In principle this might reduce the magnitude of $A_N$ and in some cases
prevent a good description of data. This is particularly true for
$\pi^-$ production, where cancellations among different flavour contributions
may be stronger. In this case, in fact, the dominance of the
$d\to\pi^-$ over the $u\to\pi^-$ contribution in the fragmentation
process can be weakened by the 
expected dominance (at large values of $x$) of the $u$ over the $d$
PDF's of the polarized proton.

Let us also remind that the available sets of pion FF's are essentially
for the neutral pion case, coming from fits to data with no flavour/charge
separation. Therefore we will fix the parameters entering the Sivers
function by reproducing at best the neutral pion SSA data; only in a
second step, we look at $A_N$ for charged pion production.  This is
slightly different from the procedure employed in former studies
\cite{noiS}, where the used FF set were valence-like dominated 
(similar to our KKP-1 model set) and, adopting a simplified kinematics,
a simultaneous best fit of
charged and neutral pion SSA was performed.

Within these assumptions we get for the three different FF sets
discussed in Section \ref{chpion}, the following optimal sets of
parameters for the Sivers function
\bea 
N_u = +0.40 & a_u = 3.0 & b_u = 0.6 \nonumber \\ 
N_d = -1.00 & a_d = 3.0 & b_d = 0.5 \hspace*{2cm} {\rm (Kretzer)}\,,
\label{sivkre} \\ 
\nonumber \\
N_u = +0.40 & a_u = 3.6 & b_u = 0.6 \nonumber \\ 
N_d = -0.55 & a_d = 3.0 & b_d = 0.3 \hspace*{2cm} {\rm (KKP-1)} \>,
\label{sivkkp1} \\
\nonumber \\
N_u = +0.40 & a_u = 2.0 & b_u = 0.3 \nonumber\\ 
N_d = -0.90 & a_d = 2.0 & b_d = 0.2 \hspace*{2cm} {\rm (KKP-2)} \,.
\label{sivkkp2}
\eea
In all three cases the parameter $r$ entering the Gaussian $k_\perp$
dependence of the Sivers function is fixed at $r=0.7$ (see also 
Ref.~\cite{adm1}). \\
Note that we are employing the MRST01 set \cite{mrst01} for the 
unpolarized PDF's entering Eq.~(\ref{del2}) instead of the GRV94 set 
\cite{grv94}, adopted for instance in Ref.~\cite{adm1}; as a consequence, the
parameters for the Sivers function are slightly different.

\begin{figure} [!,h,b]
\epsfig{figure=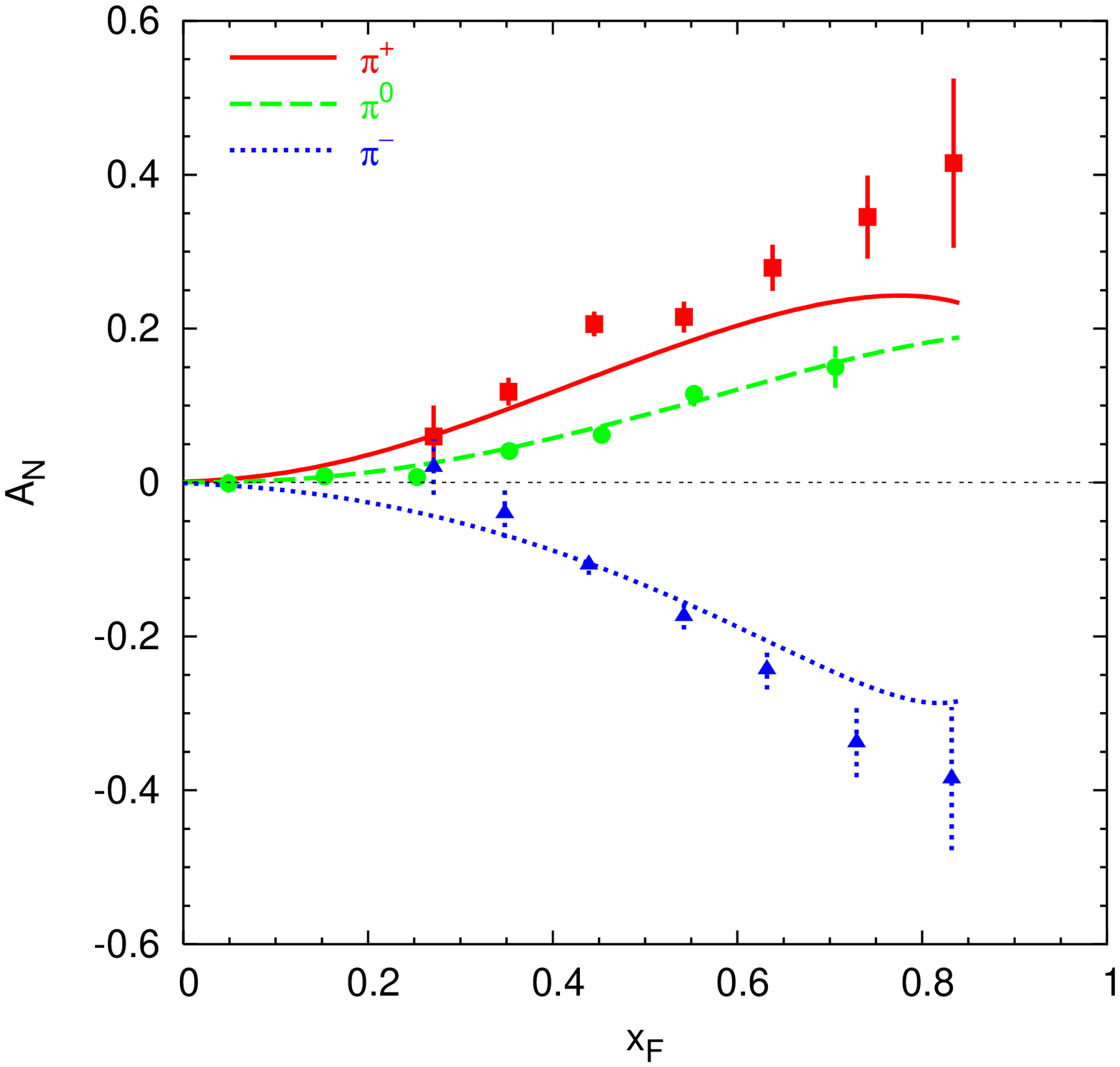,angle=-90,width=.63\textwidth}
\caption{ $A_N$ for inclusive pion production
in $pp$ collisions, at $\sqrt s = 19.4$ GeV
and fixed $p_T=1.5$ GeV/$c$, as a function of $x_F$.  The
parameterization MRST01 \cite{mrst01} for the unpolarized parton
distributions is used; fragmentation function set is from
\cite{kre}. 
For the Sivers function, see Eq.s~(\protect\ref{nqx}) and 
(\protect\ref{del2}),   
parameters are given in Eq.~(\protect\ref{sivkre}), with
$1/\beta=0.8$ GeV$/c$ and $r=0.7$. 
Data are from \cite{e704}.}
\label{ankr}
\end{figure}

\begin{figure} [!,h,t]
\epsfig{figure=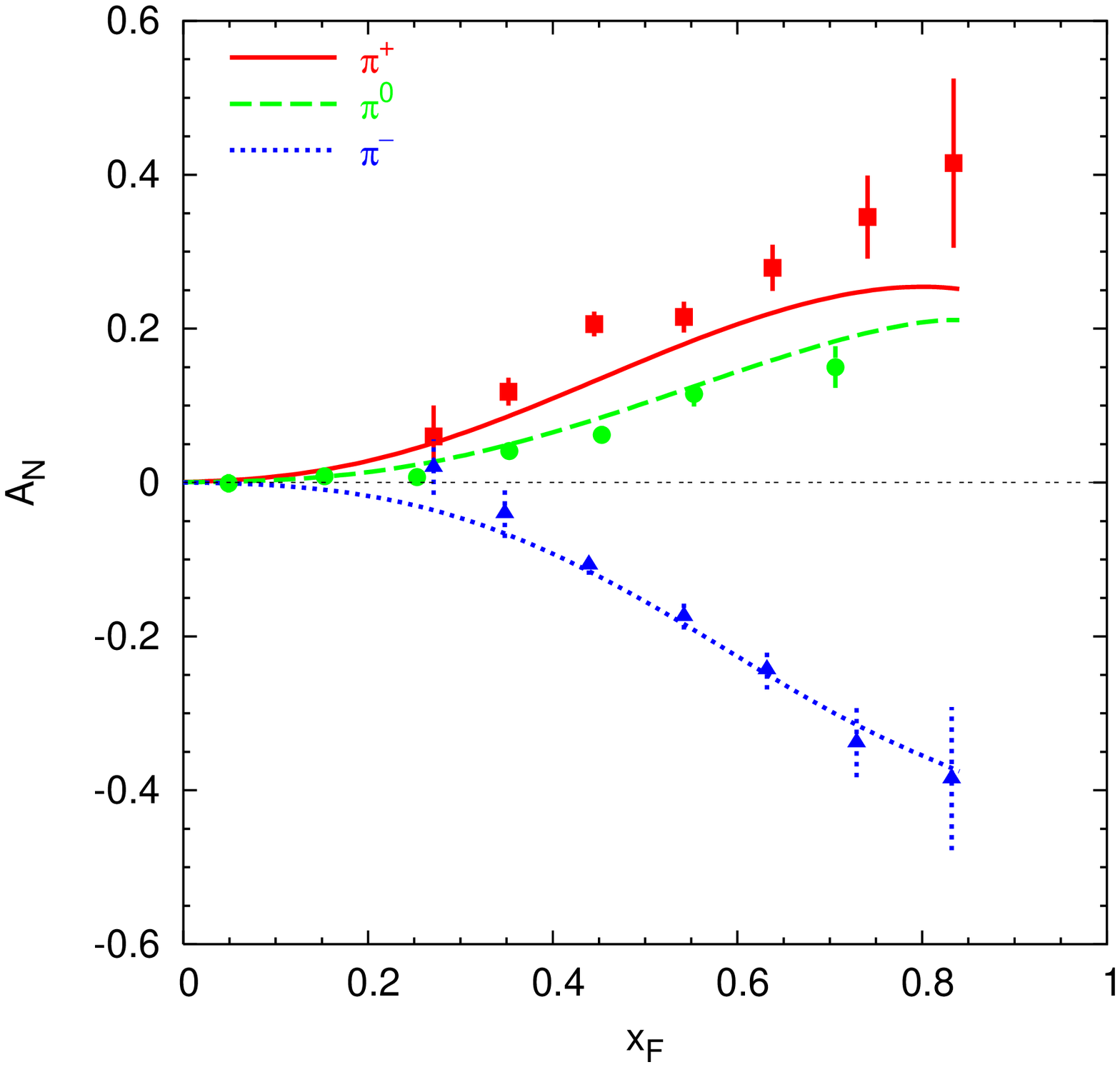,angle=-90,width=.63\textwidth}
\caption{ $A_N$ for inclusive pion production
in $pp$ collisions, at $\sqrt s = 19.4$ GeV
and fixed $p_T=1.5$ GeV/$c$, as a function of $x_F$.  The
parameterization MRST01 \cite{mrst01} for the unpolarized parton
distributions is used; fragmentation function set is KKP-1 (see
Section \ref{chpion}).
For the Sivers function, see Eq.s~(\protect\ref{nqx}) and 
(\protect\ref{del2}),   
parameters are given in Eq.~(\protect\ref{sivkkp1}), with
$1/\beta=0.8$ GeV$/c$ and $r=0.7$. 
 Data are from \cite{e704}.}
\label{an1}
\end{figure}

\begin{figure} [!,h,b]
\epsfig{figure=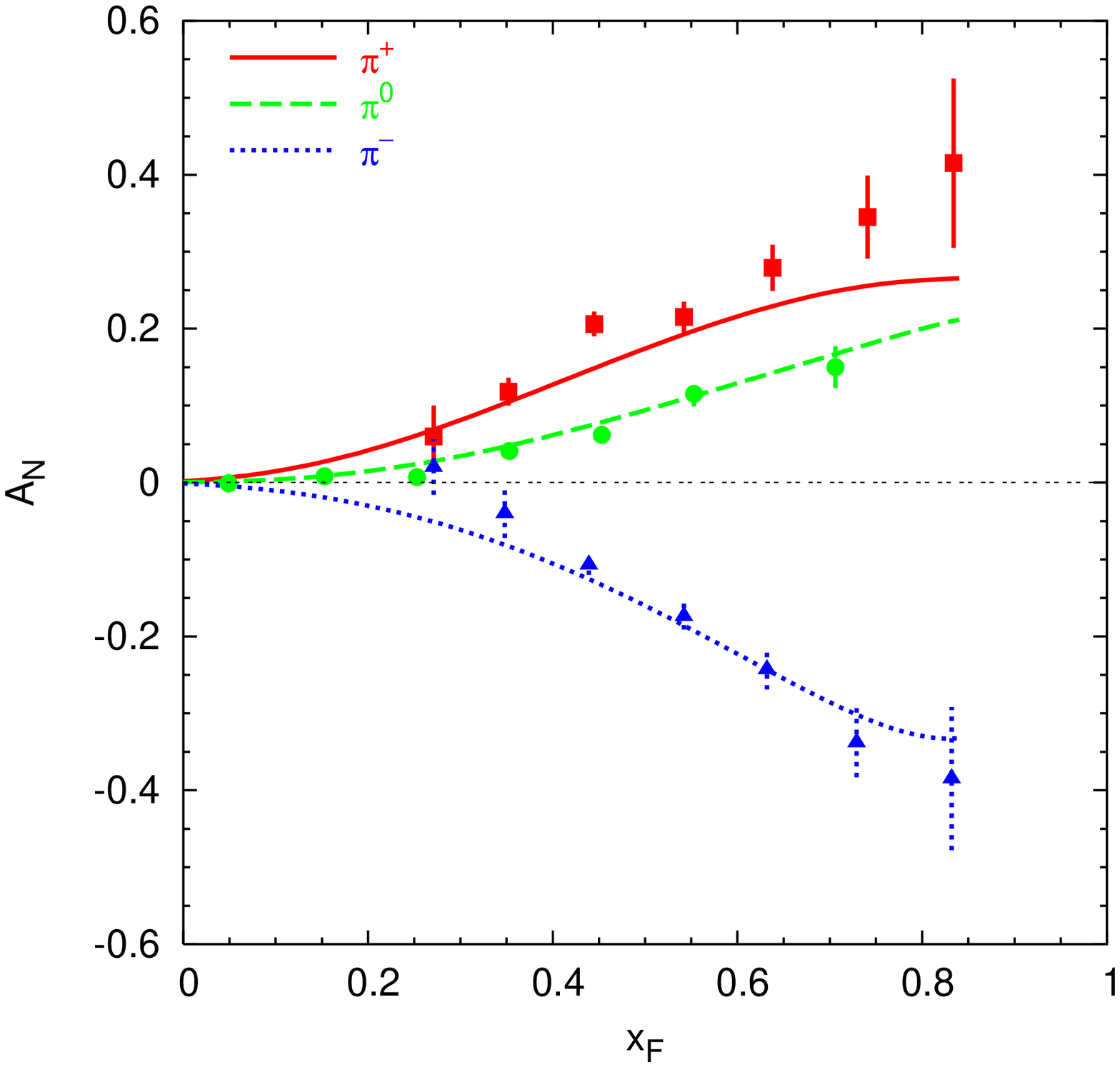,angle=-90,width=.63\textwidth}
\caption{ $A_N$ for inclusive pion production
in $pp$ collisions, at $\sqrt s = 19.4$ GeV
and fixed $p_T=1.5$ GeV/$c$, as a function of $x_F$.  The
parameterization MRST01 \cite{mrst01} for the unpolarized parton
distributions is used; fragmentation function set is KKP-2 (see
Section \ref{chpion}).
For the Sivers function, see Eq.s~(\protect\ref{nqx}) and 
(\protect\ref{del2}),   
parameters are given in Eq.~(\protect\ref{sivkkp2}), with
$1/\beta=0.8$ GeV$/c$ and $r=0.7$. 
Data are from \cite{e704}.}
\label{an2}
\end{figure}

Corresponding estimates for SSA are shown in Fig.s~\ref{ankr},
\ref{an1}, and \ref{an2}.  We first notice that using the Kretzer FF
set, where the non-leading quark FF is relevant over a large range of
$z$ values, a simultaneous good description of neutral and charged
pion SSA's seems to be more difficult (see Fig.~\ref{ankr}). In the charged
pion case, in fact, cancellations between leading and non-leading
contributions in the numerator of $A_N$ are more effective.
Even maximizing in size the $N_d$ parameter, see Eq.~(\ref{sivkre}),
but keeping the agreement with $\pi^0$ SSA data,
we are not able to give a description of the $A_N$
data as good as for the other two sets.
Apparently, the best agreement with data is obtained with the KKP-2
model set (see Fig.~\ref{an2}) where, without any need of maximizing
the $N_d$ parameter, see Eq.~(\ref{sivkkp2}), the concordance with neutral and
charged pion SSA is quite good. Concerning the KKP-1 model set, where 
non-leading quark FF's are strongly suppressed, it also allows a 
reasonable description of data (see Fig.~\ref{an1}), being a factor of 
2 far from saturating the positivity bound for both flavours,
Eq.~(\ref{sivkkp1}).
Notice that in all cases the $a$, $b$ parameters cannot be changed
very much if one wants to correctly reproduce the shape of the SSA. 
Therefore, the above conclusions can hardly be changed by modifying 
these parameters.

We have also considered a different functional form for
${\mathcal H}(k_\perp)$ in the $k_\perp$-dependent part of the Sivers 
function, Eq.s~(\ref{hk}) and (\ref{hhk}):
\be
{\mathcal H}'(k_\perp) = \frac{2\,k_\perp\,M_0}{k_\perp^2+M_0^2} \,,
\label{hhk1}
\ee
where $M_0$ is a typical hadronic scale parameter.
While still ensuring the positivity bound, Eq.~(\ref{bound2}), this form
is less (power-like) suppressed at large $k_\perp$ values as compared
with the Gaussian-like form of Eq.~(\ref{hhk}) adopted in our
calculations. We have explicitly checked that taking
$M_0 = 1/\beta= 0.8$ GeV$/c$ (a natural choice that also minimizes the
number of free parameters), and keeping the same sets of parameters as
given in Eq.s~(\ref{sivkre}), (\ref{sivkkp1}), and (\ref{sivkkp2}), we
are still able to describe the E704 data for $A_N$, at the same level
of accuracy as shown in the plots discussed above. 

Another result of this analysis, which consistently treats the full
$\bfk_\perp$ kinematics in the calculation of both the numerator and
the denominator of $A_N$, is that all the main features of the Sivers
effect and Sivers function found in previous papers \cite{noiS} based on a
simplified kinematics, are confirmed.
On the contrary, the inclusion of full
partonic kinematics seems to substantially reduce the role of the
Collins mechanism in the $p^\uparrow p\to\pi\,X$ process (see \cite{abdlm}) 

As a next step, we may now use these findings to give estimates of
$A_N$ for pion production at the much larger energies reached at the
RHIC collider, namely $\sqrt{s}=200$ GeV, but almost in the same $p_T$
region (so that the so-far unknown pQCD evolution of the Sivers function
is not relevant). In this sense the following estimates can be
considered as real predictions of our approach. Indeed they were
obtained in a preliminary version of this study \cite{dmproc}
before the new STAR data for neutral pion SSA \cite{star04} became
available. Note however that in this analysis we have not considered
the possible role of Sudakov suppression factors~\cite{suda}.

\begin{figure} [!,h,t]
\epsfig{figure=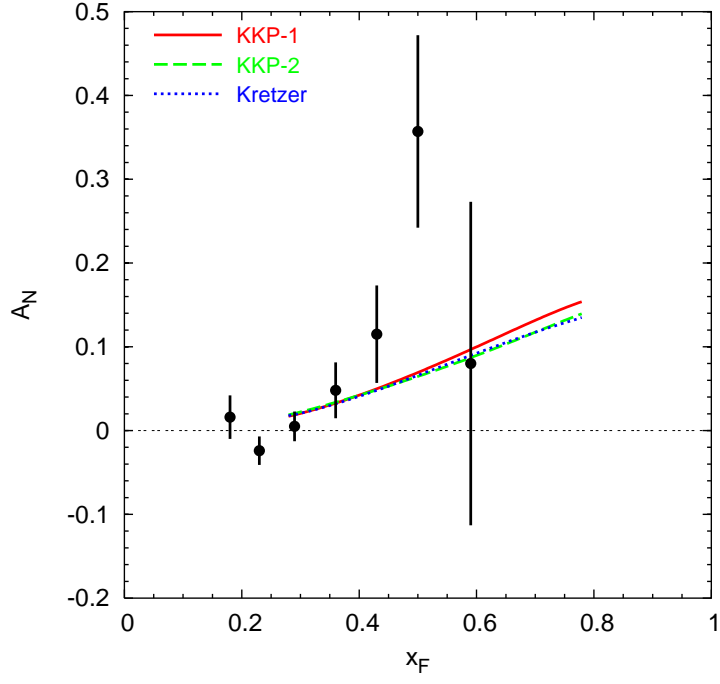,angle=-90,width=.63\textwidth}
\caption{ $A_N$ for inclusive $\pi^0$ production
in $pp$ collisions, at $\sqrt s =
200$ GeV and fixed pseudo-rapidity $\eta=3.8$, as a function of $x_F$.
The parameterization MRST01 \cite{mrst01} for the unpolarized parton
distributions is used.  Curves are for different fragmentation
function sets and corresponding Sivers function parameterizations (see
text).  Data are from \cite{star04}.}
\label{anstar}
\end{figure}

In Fig.~\ref{anstar} we show our results obtained using the three sets of
parameters for the Sivers function discussed above and compare them
with STAR experimental data \cite{star04}.  Curves are calculated at
fixed pseudo-rapidity $\eta=3.8$, so that $p_T$ and $x_F$ are
correlated. To be safe with our pQCD approach, we consider only the
region of $p_T>$1-1.5 GeV$/c$; our curves are then cut at $p_T$ values
around 1.25 GeV$/c$ (which implies $x_F\agt0.3$).  This comparison
indicates  reasonable agreement with data, although error bars are
still quite big at large $x_F$. We stress the fact that, as predicted
theoretically, sizable pion SSA, with features similar to those found
in the E704 kinematical range, are confirmed experimentally
even at such large c.m. energies.

\subsection{ Sivers effect in $p^\uparrow p\to\gamma X$}
\label{ssag}

Direct photon production in singly polarized $pp$ collisions is
definitely of interest in this context, since it allows
(unambiguous) access to the Sivers effect.
In this case there is clearly no contamination from fragmentation
processes (Collins effect); in the positive, moderately large $x_F$
region only the Sivers effect could be responsible for such an
asymmetry in our approach.
Therefore, the numerator of the SSA, Eq.~(\ref{an}), is given by an
expression similar to that for the unpolarized cross section,
Eq.~(\ref{ppg}), substituting the unpolarized PDF
$\hat{f}_{a/p}(x_a,\bm{k}_{\perp a})$ with the corresponding Sivers
distribution  $\Delta^{\!N}{f}_{a/p^\uparrow}(x_a,\bm{k}_{\perp a})$;
the denominator is simply given by twice the unpolarized cross section. 

Unfortunately the few available data, at very small
$x_F$, are compatible with zero within the (large) errors.  Therefore
a comparison with our predictions is somehow inconclusive.  No test of
our assumption of a valence-quark dominated Sivers function is
possible either.

\begin{figure} [!,h,b]
\epsfig{figure=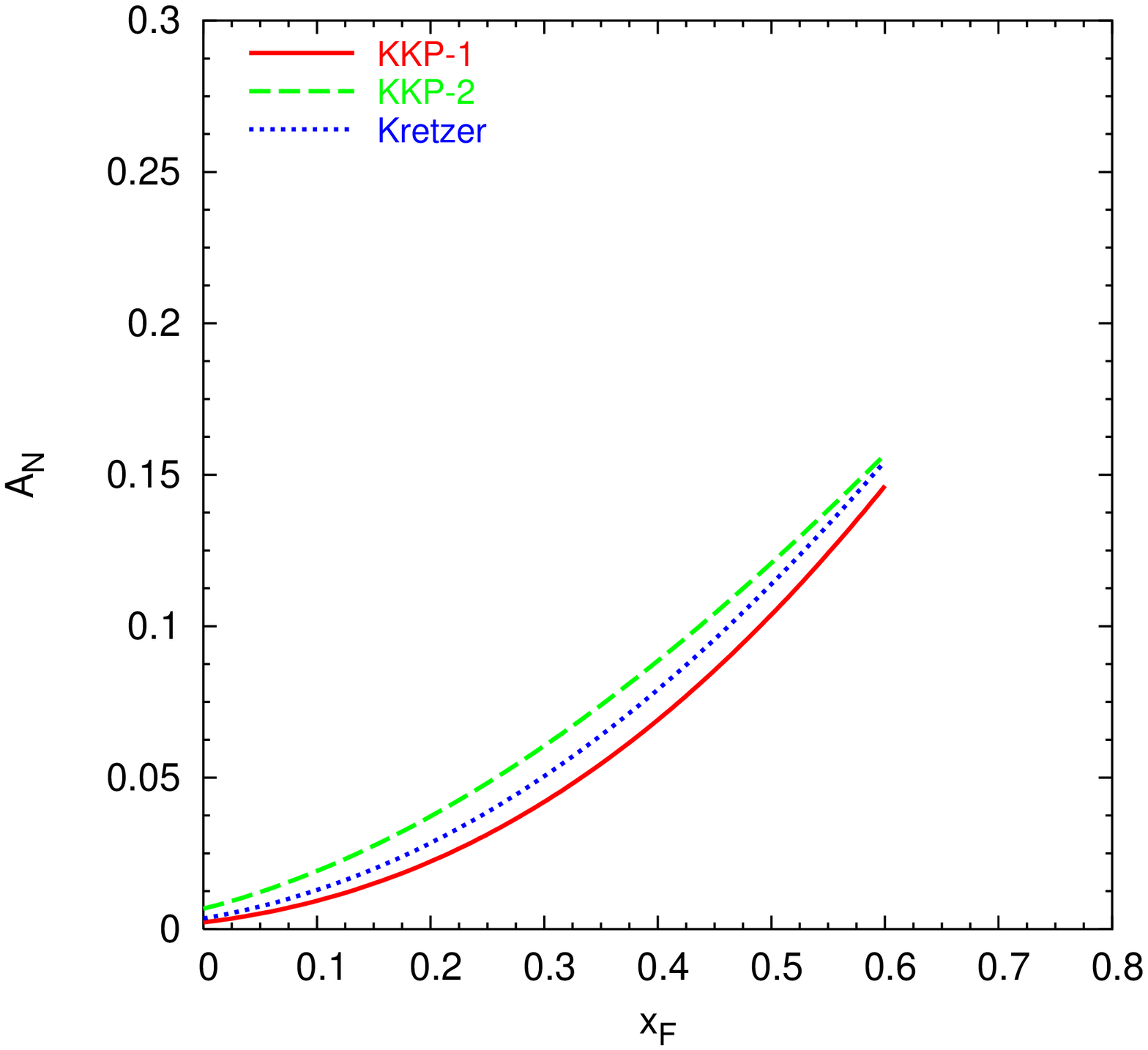,angle=-90,width=.63\textwidth}
\caption{ $A_N$ for inclusive photon production
in $pp$ collisions, at $\sqrt s = 19.4$
GeV and fixed $p_T=2.7$ GeV$/c$, as a function of $x_F$.  The
parameterization MRST01 \cite{mrst01} for the unpolarized parton
distributions is used.  Curves correspond to different Sivers function
parameterization sets (see text).   
}
\label{ange704}
\end{figure}

In Fig.~\ref{ange704} we show our predictions, utilizing the three
different Sivers function parameterizations considered,
for the kinematics of the E704 experiment, that is at
$\sqrt s = 19.4$ GeV and fixed $p_T= 2.7$ GeV/$c$, as a function of $x_F$.
The calculated asymmetry is sizable at medium-large values of $x_F$,
and 
could be measurable in principle. Experimental data at intermediate
$x_F$ values would be of great help.

\begin{figure} [!,h,t]
\epsfig{figure=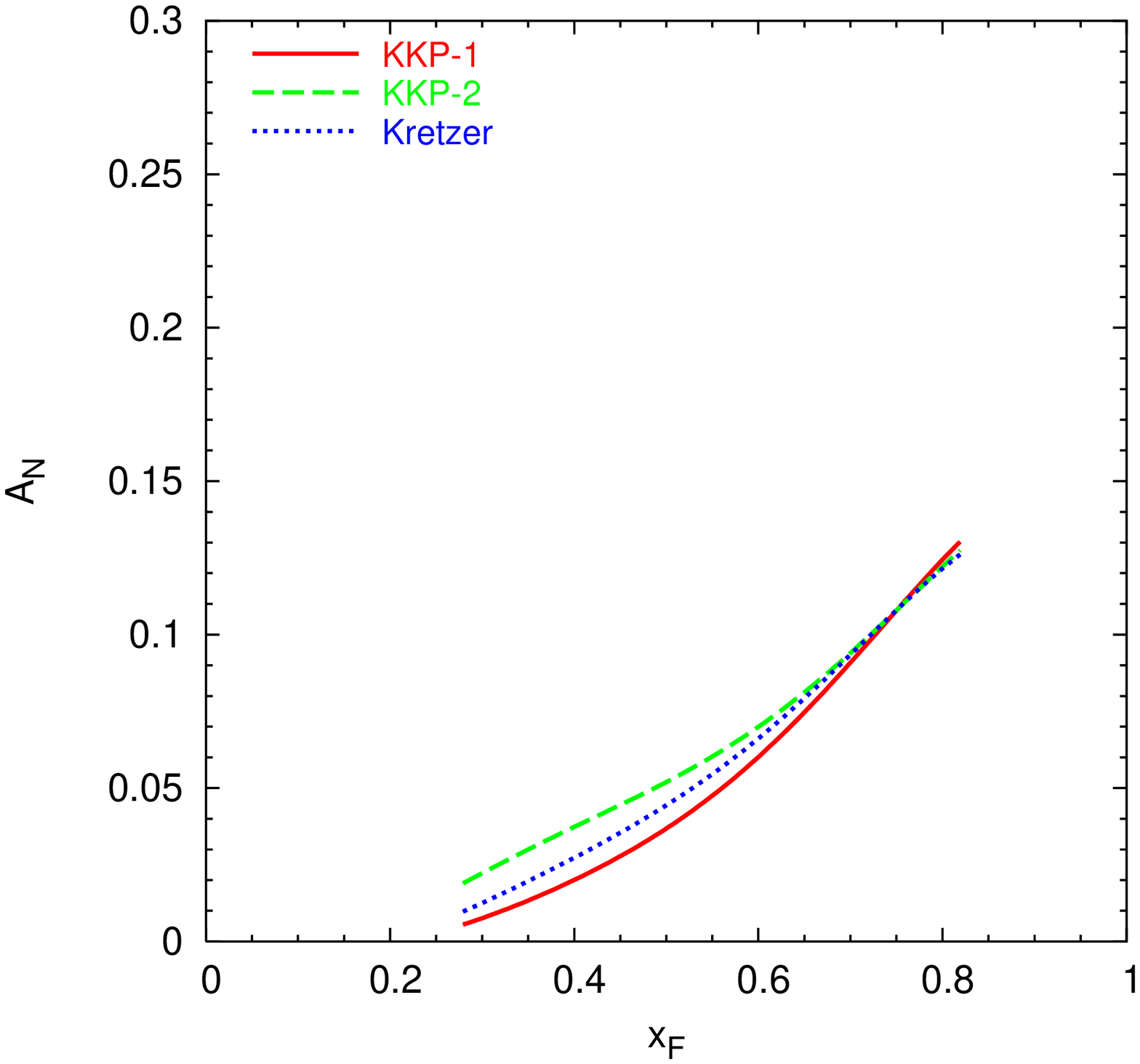,angle=-90,width=.63\textwidth}
\caption{ $A_N$ for inclusive photon production
in $pp$ collisions, at $\sqrt s = 200$
GeV and fixed rapidity $y=3.8$, as a function of $x_F$.  The
parameterization MRST01 \cite{mrst01} for the unpolarized parton
distributions is used.  Curves correspond to different Sivers function
parameterization sets (see text).   
}
\label{angstar}
\end{figure}

We also show our results for STAR kinematics in Fig.~\ref{angstar},
namely at $\sqrt s =200$ GeV and fixed rapidity $y=3.8$: once again
sizable SSA are found which could be checked experimentally.

We can then conclude that SSA in direct photon production could be an
important tool in understanding the Sivers mechanism.  The same is true for
any process where only one at a time of all possible effects might be at
work. \\ 
This is the case for instance of the single polarized Drell-Yan
process, where by suitably integrating on the angular variables 
identifying the leptonic scattering plane only the Sivers effect would survive
\cite{adm1}; another key process could be heavy meson production in
hadronic collisions, where the dominance of quark and gluon
annihilation channels can enhance or even select out the Sivers
effect \cite{noiD}.

\section{Conclusions}
\label{concl}

In this paper we have studied the role of partonic intrinsic
transverse momentum in inclusive particle production in (un)polarized 
high-energy hadronic reactions. Several papers in the framework of LO 
and NLO collinear pQCD have already been devoted to this subject in
recent years.
Our analysis is different in that it is mainly aimed at stressing the 
connection with recent approaches to the problem of the large single 
spin asymmetries measured in inclusive pion production at different
c.m.~energies at large positive $x_F$ and moderately large $p_T$.
A lot of theoretical work has shown that LO pQCD with the inclusion of
partonic intrinsic transverse momentum and a new class of leading twist,
spin and $\bm{k}_\perp$-dependent PDF's and FF's is a promising
approach to this problem.
However, in those papers it was never shown in detail to what extent this
approach is able to reproduce unpolarized cross sections for the same
processes in similar kinematical conditions.
Therefore, in the first part of this paper  we have addressed this
problem in a systematic way, with a full account of $\bm{k}_\perp$ effects
both in the PDF's/FF's and in the elementary partonic process.
We have applied our approach to the Drell-Yan process and to inclusive 
pion and direct photon production in hadronic collisions in several 
kinematical situations of interest, performing a detailed comparison 
with available experimental data.
Our results show that, with few noticeable exceptions, our LO
approach (when complemented with proper NLO $K$-factors) 
is in reasonable agreement with a large set of 
experimental data for unpolarized cross sections.
Therefore, this gives support to the validity of the same approach in 
the study of single spin asymmetries. 

In the second part of the paper, we have then applied our approach
to the study of SSA in inclusive pion and direct photon production,
with the inclusion of Sivers contribution alone (different contributions,
like the Collins effect, in these or other processes have been or will
be studied elsewhere). We have shown that the E704 experimental results 
on pion SSA can be reproduced with good accuracy by using physically 
reasonable parameterizations of the Sivers function for valence
quark contributions (therefore, neglecting sea-quark and gluon
Sivers functions in first approximation).

These same parameterizations lead to predictions in agreement with measurements
performed by the STAR Collaboration at the RHIC-BNL accelerator at
much higher c.m. energies and similar $x_F$ and $p_T$ ranges.

We have also shown that our analysis of SSA in terms of the Sivers
effect and with a full treatment of
$\bm{k}_\perp$ effects is in good qualitative agreement and confirms all
main results and conclusions of former studies performed keeping
only leading contributions in $\bm{k}_\perp$ and using a simplified 
partonic configuration.

However, some additional comments and some words of caution, 
regarding our results on unpolarized cross sections, are in order.
The use of LO and NLO pQCD calculations (already in the collinear 
configuration) in the study of relatively low $p_T$ data faces a number
of formal problems and leads to some
model dependent choices regarding e.g. the factorization scale and the
strong coupling constant behaviour. Moreover, from the numerical point 
of view, the lower the $p_T$ values considered, the more relevant
are these model dependences.

Our results, which cover a range of $p_T$ values among $1$--$15$ GeV/$c$
for the cross sections and of a few GeV$/c$ at most for the SSA,
have been obtained
by performing a number of choices that we believe very reasonable and 
are largely adopted in the literature.
Admittedly, much work
remains to be done, in particular on the more
formal aspects of the approach.
We can however conclude with good confidence that
there is not presently  any strong argument or evidence 
against the applicability of our approach to SSA calculations coming from
unpolarized cross section results, in particular for the kinematical
situations of relevance for SSA.
Within the uncertainties inherent the pQCD approach discussed above
(hopefully less relevant for the SSA which are expressed as
ratios of cross sections) there is good agreement among our results and
experimental data.
Concerning those few cases where our approach clearly fails to
reproduce the results, we want to notice the following: 1) We have 
considered a very large set of data in different kinematical
situations and from both fixed target and collider experiments; 
2) Even NLO pQCD in collinear configuration faces several problems in fitting 
simultaneously the large collection of data considered here.

Let us finally add some words concerning transverse hyperon polarization
in unpolarized hadronic collisions. It was shown that our approach is in
principle able to reproduce most of the striking features of the 
experimental data available; the role of the so-called polarizing 
fragmentation function was emphasized. However, a combined analysis
of SSA and unpolarized cross sections, like that performed here for
the Drell-Yan process and inclusive pion, photon productions, is still
lacking. This problem deserves a separate analysis which
is under way and will be 
presented elsewhere.

\acknowledgments

We are grateful to M. Anselmino and E. Leader for a critical reading
of the manuscript and for useful comments. A special thank to W. Vogelsang
who helped us in understanding the role of NLO corrections.  We
acknowledge partial support by M.I.U.R. (Ministero dell'Istruzione,
dell'Universit\`a e della Ricerca) under Cofinanziamento P.R.I.N. 2003.

\appendix

\section{Full $\bfk_\perp$ kinematics}

We give here a detailed treatment of partonic kinematics with proper
inclusion of transverse momentum effects, along the same lines of
\cite{cont}.  Let us consider the hadronic reaction $ A B \to C X$ in
the $A$-$B$ center of mass frame with $A$ moving along the
positive $Z$ axis and fix the scattering plane as the $X$-$Z$ plane.
The 4-momenta of hadrons $A,B,C$ read 
\be 
p_A^\mu = (E_A,0,0,P)
\;\;\;\; p_B^\mu = (E_B,0,0,-P) \;\;\;\; p_C^\mu = (E_C,p_T,0,p_L) \,,
\ee 
with $E_{A,B} = \sqrt{P^2+m_{A,B}^2} $ and $E_C=
\sqrt{p_T^2+p_L^2+m_C^2}$.  For equal-mass initial hadrons
($m_A=m_B=m$), we have $E_A = E_B= \sqrt s/2$, with $s=(p_A+p_B)^2$.

For massless partons $a,b$ inside hadrons $A,B$ we introduce
light-cone momentum fractions $x_a=p_a^+/p_A^+$ , $x_b=p_b^-/p_B^-$
and transverse momenta $\bfk_{\perp a}$, $\bfk_{\perp b}$. 
Their four-momenta then read 
\bea
\label{papb}
p_a^\mu & = & x_a w \frac{\sqrt s}{2} \left( 1 + \frac{k_{\perp a}^2}{x_a^2
w^2 s}, \frac{2k_{\perp a}}{x_a w \sqrt s} \cos\phi_a, 
\frac{2k_{\perp a}}{x_a w \sqrt s} 
\sin\phi_a, 1 - \frac{k_{\perp a}^2}{x_a^2 w^2 s} \right) \,,\nonumber\\
p_b^\mu & = & x_b w \frac{\sqrt s}{2} \left( 1 + \frac{k_{\perp b}^2}{x_b^2
w^2 s}, \frac{2k_{\perp b}}{x_b w \sqrt s} \cos\phi_b, 
\frac{2k_{\perp b}}{x_b w \sqrt s} 
\sin\phi_b, - 1 + \frac{k_{\perp b}^2}{x_b^2 w^2 s} \right) \,, 
\eea 
where the factor $w$, 
defined as $w = \left[1+\sqrt{1-4m^2/s}\right]/2$, can be safely
taken equal to one for high-energy processes, as considered in this
paper.  In Eq.~(\ref{papb}) $k_{\perp a,b} = |\bfk_{\perp a,b}|$,  
and $\phi_{a,b}$
are the azimuthal angles of parton $a,b$ three-momenta in the hadronic
c.m. frame.

The four-momentum of fragmenting parton $c$ is given in terms of the
observed hadron momentum $p_C^\mu$, of the light-cone momentum
fraction $z=p_C^+/p_c^+$ and of the transverse momentum of hadron $C$ 
with respect to parton $c$ light-cone direction, $\bfk_{\perp C}$. 
In order to have all
transverse momenta defined in the hadronic c.m.~frame, we define the 
(two-dimensional) transverse momentum, $\bfk_{\perp C}$, 
as a genuine three-momentum: 
\be
\label{kc}
\bfk_C = k_C (\sin\theta_{k_C}\cos\phi_{k_C},
\sin\theta_{k_C}\sin\phi_{k_C}, \cos\theta_{k_C} ) \,, 
\ee 
and impose the orthogonality condition $\bfk_C\cdot\bfp_c = 0$, see below.
Keeping hadron and parton masses into account, the parton
four-momentum, $p_c^\mu = (E_c,\bfp_c)$, is then given as 
\bea
\label{pcg}
\bfp_c & = & \frac{\sqrt{E_c^2-m_c^2}}{\sqrt{\bm{p}_C^2-k_C^2}}\,
(\bfp_C-\bfk_C) = \frac{\sqrt{E_c^2-m_c^2}}{\sqrt{\bm{p}_C^2-k_C^2}} \,
(p_T - k_C\sin\theta_{k_C}\cos\phi_{k_C}, -
k_C\sin\theta_{k_C}\sin\phi_{k_C}, p_L-k_C\cos\theta_{k_C} )\,, \\
\label{Ecg}
E_c & = & \frac{E_C + \sqrt{\bm{p}_C^2-k_C^2}}{2z} \left[ 1 +
 \frac{z^2m_c^2}{\left( E_C+\sqrt{\bm{p}_C^2-k_C^2}\right)^2} \right]\>.  
\eea
When, as it is the case in this paper, hadron $C$ is a pion and only
light quark and gluon fragmentation is considered, hadron and parton
masses can safely be neglected in the kinematics; eqs.~(\ref{pcg}) and
(\ref{Ecg}) then simplify to 
\bea
\label{pc}
\bfp_c 
& = &
 \frac{E_c}{\sqrt{E_C^2-k_C^2}} \, 
(p_T - k_C\sin\theta_{k_C}\cos\phi_{k_C}, -
 k_C\sin\theta_{k_C}\sin\phi_{k_C}, p_L-k_C\cos\theta_{k_C} ) \,,\\
\label{Ec}
E_c & = & \frac{E_C + \sqrt{E_C^2-k_C^2}}{2z} \>.  
\eea
On the other hand for heavy hadrons (like $\Lambda$ hyperons) and/or heavy
quarks (like charm quark fragmentation into $D$ mesons), mass
corrections can be comparable with $\bfk_\perp$ effects.

Another ingredient entering our basic factorization formula,
Eq.~(\ref{kqcd}), is the Jacobian factor
$J(z,|\bfk_C|)$
connecting the parton $c$ to
hadron $C$ invariant phase space, defined as 
\beq
\frac{d^3\bm{p}_c}{E_c} =
\frac{1}{z^2}\, J(z,|\bfk_C|)\frac{d^3\bm{p}_C}{E_C} \,, 
\eeq
which for collinear and massless particles reduces simply to 
$J = 1$.  In the most general case, after some algebra, one gets 
\beq
\label{jac}
J(z,|\bfk_C|) =
\frac{\left(E_C+\sqrt{\bm{p}_C^2-k_C^2}\right)^2}{4(\bm{p}_C^2-k_C^2)} \,
\left[1- \frac{z^2 m_c^2}{\left(E_C+\sqrt{\bm{p}_C^2-k_C^2}\right)^2}
\right]^2 \Rightarrow
\frac{\left(E_C+\sqrt{E_C^2-k_C^2}\right)^2}{4(E_C^2-k_C^2)}\,,
\eeq 
where the last expression holds for massless partons and hadrons.

The orthogonality condition $\bfk_C\cdot \bfp_c=0$ mentioned above is
realized through the following relation 
\beq
\label{kcpc}
 d^2\bfk_{\perp C} =  d^3\bfk_C \, \delta(\bfk_C\cdot
\hat{\bfp}_c) \,, 
\eeq 
where $\hat{\bfp}_c$ is the unit vector along
the direction of motion of parton $c$.  By using Eqs.~(\ref{kc}) and
(\ref{pcg}), Eq.~(\ref{kcpc}) becomes 
\bea
\label{dephik}
 d^2\bfk_{\perp C} & = & k_C dk_C\, d\theta_{k_C} \, d\phi_{k_C} \,
\frac{\sqrt{\bm{p}_C^2-k_C^2}}{p_T \sin\phi^0_{k_C}}\, \left[
\delta(\phi_{k_C} - \phi^0_{k_C}) + \delta(\phi_{k_C} - (2\pi -
\phi^0_{k_C}))\right] \,,\\
\label{phik}
\cos\phi^0_{k_C} & = &
\frac{k_C-p_L\cos\theta_{k_C}}{p_T\sin\theta_{k_C}}, \;\; 0 \le
\phi^0_{k_C} \le \pi \>.  
\eea 
In this way the integration over
$\phi_{k_C}$ can be carried out directly (notice that there are two
possible solutions to be considered).

With the expression of parton momenta given in Eqs.~(\ref{papb}) and
(\ref{pc}) one can calculate the partonic Mandelstam invariants; by
exploiting the elastic, massless parton constraint $\hat s +
\hat t + \hat u=0$, one can in turn fix the value of $z$:
\bea
\label{shat}
\hat s = (p_a+p_b)^2 & = & x_a x_b s \left[1 - 2 \frac{k_{\perp a}
    k_{\perp b}}{x_ax_b s} \cos(\phi_a-\phi_b) + 
\frac{k_{\perp a}^2 k_{\perp b}^2}{x_a^2 x_b^2 s^2}
\right] \,,\\ 
\hat t = (p_a - p_c)^2 & = & \frac{T}{z} \,,\\ 
\hat u = (p_b - p_c)^2 & = & \frac{U}{z} \,,\\
\label{xc}
\hat s\, \delta(\hat s + \hat t + \hat u) & = & z\, \delta\left(z
+ \frac{T+U}{\hat s}\right) \,, 
\eea 
where the two functions $T$ and
$U$, easily obtained from the explicit expressions of parton momenta,
are independent of $z$.  For heavy quarks, one
can still fix $z$ from the elastic constraint but in this case 
the expressions for $\hat t$ and $\hat u$, in terms of $z$, are more involved.

We now discuss, limiting to the massless case, 
the constraints on the full phase space entering our
factorized expression for (un)polarized cross sections.  Besides the
trivial bounds $0 < x_{a,b}, z < 1$, $0 \le \phi_{a,b} \le 2\pi$ and $0
\le \theta_{k_C} \le \pi$, we require that, even including intrinsic
transverse momentum effects, $a)$ each parton keeps moving along the
same direction as its parent hadron, $ \bfp_{a(b)} \cdot \bfP_{A(B)} >
0$, and $b)$ the parton energy is not larger than the parent hadron
energy, $E_{a(b)} \le E_{A(B)}$.  This implies the following bounds
\beq
\label{kakblim}
k_{\perp a(b)}/\sqrt s < \min \left[x_{a(b)}, \sqrt{x_{a(b)}(1-x_{a(b)})}
\right]\>.  
\eeq 
Analogously, for the fragmentation process $c\to C+X$
we require $\bfp_c\cdot \bfP_C> 0$ and $E_C \le E_c$ (both fulfilled
by Eq.~(\ref{Ec}), where we have consistently disregarded the solution
$E_c=\left[E_C - \sqrt{E_C^2-k_C^2}\right]/(2z)$).  The last constraint implies
the following bound on $k_C$, at fixed $z$ 
\beq
\label{kclim}
k_C/ E_C \le 1\;\, (z \le 1/2);\;\;\;\; k_C/E_C \le
2\sqrt{z(1-z)}\;\, (z > 1/2)\>.  
\eeq
By requiring $|\cos\phi^0_{k_C}|\leq 1$, see Eq.~(\ref{phik}), we have a
further constraint on $k_C$, at fixed $\theta_{k_C}$, namely 
\beq
\label{kclim2}
p_L\cos\theta_{k_C} - p_T\sin\theta_{k_C} \le k_C \le
p_L\cos\theta_{k_C} + p_T\sin\theta_{k_C}\>.  
\eeq
A word of caution is needed for the partonic flux factor when incoming
partons are not collinear. As shown in \cite{cahn} the correct
convolution formula connecting the partonic to the hadronic process is
expressed in terms of partonic and hadronic tensors, that is in terms
of scattering amplitudes squared rather than directly in terms of
cross sections.  The convolution then only involves parton-hadron
light-cone energy ratios and no partonic flux, which would depend
on the relative azimuthal angle between parton momenta, enters.

More precisely, for the reaction $AB\to c d$ one has 
\bea 
|M|^2_{AB\to cd} & = & \int dx_a d^2\bm{k}_{\perp a} dx_b d^2\bm{k}_{\perp  b}
\, \hat{f}_{a/A}(x_a,\bfk_{\perp a})
\hat{f}_{b/B}(x_b,\bfk_{\perp b})\, |M|^2_{ab\to cd}\, 
\frac{P_A^+ P_B^+}{p_a^+ p_b^+} \nonumber\\
\label{amp} 
 & = & \int dx_a d^2\bm{k}_{\perp a} dx_b d^2\bm{k}_{\perp b}
 \hat{f}_{a/A}(x_a,\bfk_{\perp a}) \hat{f}_{b/B}(x_b,\bfk_{\perp b})
 \,|M|^2_{ab\to cd}\,
 \frac{1}{x_a x_b} \>.  
\eea 
In order to obtain the proper normalized $AB\to cd$ cross section 
one has to divide Eq.~(\ref{amp}) by the {\it hadronic} flux ($2s$). 
On the other hand since our convolution formula is given in terms of
partonic cross sections, which are normalized to the partonic flux
$2\hat s$, to restore proper normalization 
we have to rescale them by the factor $\hat s/(x_ax_b s)$.


\newpage

\newpage

\newpage

\newpage

\newpage

\newpage

\newpage

\newpage

\newpage

\newpage

\begin{thebibliography}{}
\bibitem{fff}   R.D. Field and R.P. Feynman, \PR {D15}, 2590 (1977);
                R.P. Feynman, R.D. Field and G.C. Fox,
                \PR {D18}, 3320 (1978).
\bibitem{pkt}   X.-N. Wang, \PR {C61}, 064910 (2000);
                Y. Zhang, G. Fai, G. Papp, G. Barnaf\"oldi, and
                P. L\'evai, \PR {C65}, 034903 (2002).
\bibitem{ww}    C.Y Wong, and H. Wang, \PR {C58}, 376 (1998).
\bibitem{apa98} L. Apanasevich, {\em et al.} (E706 Collab.), 
                \PRL {81}, 2642 (1998).
\bibitem{kane}  G.L. Kane, J. Pumplin, and W. Repko,
                \PRL {25}, 1689 (1978). 
\bibitem{siv}   D. Sivers, \PR {D41}, 83 (1990); \PR {D43}, 261
                (1991).
\bibitem{noiS}  M. Anselmino, M. Boglione, and F. Murgia,
                \PL {B362}, 164 (1995);
                M. Anselmino, and F. Murgia, \PL {B442}, 470 (1998).
\bibitem{newsiv}S.J. Brodsky, D.S. Hwang, and I. Schmidt,
                \PL {B530}, 99 (2002); \NP {B642}, 344 (2002);
                J.C. Collins, \PL {B536}, 43 (2002).
\bibitem{bs03}  C. Bourrely and J. Soffer,
                e-Print Archive: hep-ph/0311110.
\bibitem{noiC}  M. Anselmino, M. Boglione, and F. Murgia,
                \PR {D60}, 054027 (1999);
                M. Anselmino, M. Boglione, J. Hansson, and F. Murgia,
                \EPJ {C13}, 519 (2000);
                M. Boglione, and E. Leader, \PR {D61}, 114001 (2000).
\bibitem{col}   J.C. Collins, \NP {B396}, 16 (1993).
\bibitem{css}   J.C. Collins, D.E. Soper and G. Sterman, \NP {B250},
                199 (1985); J.C. Collins and D.E. Soper, \NP {B193},
                381 (1981).
\bibitem{jimay} X. Ji, J. Ma, F. Yuan,
                e-Print Archive: hep-ph/0404183; hep-ph/0405085.
\bibitem{muld}  P.J. Mulders, and R.D. Tangerman, \NP {B461}, 197
                (1996), erratum ibid.~{\bf B484}, 538 (1997); 
		D. Boer, and P.J. Mulders, \PR {D57}, 5780 (1998).
\bibitem{boer}  D. Boer, \PR {D60}, 014012 (1999).
\bibitem{abdlmm}M. Anselmino, M. Boglione, U. D'Alesio, E. Leader,
                S. Melis, and F. Murgia, in progress.
\bibitem{abdm1} M. Anselmino, D. Boer, U. D'Alesio, and F. Murgia,
               \PR {D63}, 054029 (2001); \PR {D65}, 114014 (2002).
\bibitem{enzo}  V. Barone, A. Drago, and P.G. Ratcliffe, 
                Phys.\ Rep.\ {\bf 359}, 1 (2002). 
\bibitem{abdlm} M. Anselmino, M. Boglione, U. D'Alesio, E. Leader, and
                F. Murgia, in progress.
\bibitem{adm1}  M. Anselmino, U. D'Alesio, and F. Murgia,
                \PR {D67}, 074010 (2003).
\bibitem{ss}    D.V. Shirkov and I.L. Solovtsov, \PRL {79}, 1209 (1997).
\bibitem{aue}   P. Aurenche, R. Baier, M. Fontannaz, and D. Schiff,
                \NP {B297}, 661 (1988).
\bibitem{ave}   F. Aversa, P. Chiappetta, M. Greco, and J.Ph. Guillet,
                \NP {B327}, 109 (1989).
\bibitem{wer}   W. Vogelsang, private communication.
\bibitem{mrst01}A.D. Martin, R.G. Roberts, W.J. Stirling, and R.S. Thorne,
                \PL {B531}, 216 (2002).
\bibitem{app}   G. Altarelli, G. Parisi and R. Pretonzio,
                \PL {B76}, 351 (1978); 356 (1978); 
                H. Fritzch and P. Minkowski, \PL {B73}, 80 (1978).
\bibitem{ito}   A.S. Ito, {\em et al.}, \PR {D23}, 604 (1981).
\bibitem{mor91} G. Moreno, {\em et al.}, \PR {D43}, 2815 (1991).
\bibitem{isrdy} D. Antreasyan, {\em et al.} \PRL {48}, 302 (1982).
\bibitem{vw}    W. Vogelsang and M.R. Walley, J. Phys. {\bf G23},
                Suppl.~7A, A1 (1997).
\bibitem{ow87}  J.F. Owens, Rev. Mod. Phys. {\bf 59}, 465 (1987).
\bibitem{cteq-g}J. Huston, E. Kovacs, S. Kuhlmann, H.L. Lai, 
                J.F. Owens, and W.K. Tung, \PR {D51}, 6139 (1995).
\bibitem{ll}    H.L. Lai and Hsiang-nan Li, \PR {D58}, 114020 (1998).
\bibitem{dmm}   U. D'Alesio, S. Melis, and F. Murgia in preparation.
\bibitem{e704g} D.L. Adams, {\em et al.} (E704 Collab.),
                \PL {B345}, 569 (1995).
\bibitem{wa70}  M. Bonesini, {\em et al.}, \ZP {C38}, 371 (1988).
\bibitem{isrg}  E. Anassontzis {\em et al.} (R806/807 Collab.),
                \ZP {C13}, 277 (1982).
\bibitem{apa03} L. Apanasevich, {\em et al.} (E706 Collab.), 
                \PR {D68}, 052001 (2003).
\bibitem{kee}   S. Kretzer, E. Leader, and E. Christova,
                \EPJ {C22}, 269 (2001).
\bibitem{kkp}   B.A. Kniehl, G. Kramer, and B. P\"{o}tter,
                \NP {B582}, 514 (2000).
\bibitem{isrpi} D.L. Owen {\em et al.}, \PRL {45}, 89 (1980).
\bibitem{dona}  G. Donaldson, {\em et al.}, \PL {B73}, 375 (1978).
\bibitem{adl03} S.S. Adler {\em et al.} (PHENIX Collab.),
                \PRL {91}, 241803 (2003).
\bibitem{ada96} D.L. Adams, {\em et al.} (E704 Collab.),
                \PR {D53}, 4747 (1996).
\bibitem{e704}  D.L. Adams, {\em et al.} (E704 Collab.), \PL {B345},
                569 (1995);  {\bf B261}, 197 (1991);  {\bf B264}, 462 (1991).
\bibitem{ada03} J. Adams, {\em et al.} (STAR Collab.),
                \PRL {92}, 171801 (2004).
\bibitem{kre}   S. Kretzer, \PR {D62}, 054001 (2000).
\bibitem{FNALpm}D. Antreasyan, {\em et al.}, \PR {D19}, 764 (1979).
\bibitem{ell}   E. Leader, e-Print Archive: hep-ph/0405284, 
                and private communication.
\bibitem{rodr}  P.J. Mulders and J. Rodrigues, \PR {D63}, 094021 (2001).
\bibitem{noiD}  M. Anselmino, M. Boglione, U. D'Alesio, E. Leader, and
                F. Murgia, e-Print Archive:  hep-ph/0407100.
\bibitem{grv94} M. Gluck, E. Reya, and A. Vogt, \ZP {C67}, 433 (1995).
\bibitem{dmproc}U. D'Alesio and F. Murgia,
                AIP Conf. Proc. {\bf 675}, 469 (2003);
\bibitem{star04}J. Adams, {\em et al.} (STAR Collab.),
                \PRL {92}, 171801 (2004).
\bibitem{suda}  D. Boer, \NP {B603}, 195 (2001).
\bibitem{cont}  A.P. Contogouris, R. Gaskell, and S. Papadopoulos,
                \PR{D17}, 2314 (1978);
\bibitem{cahn}  R.N. Cahn, \PR{D40}, 3107 (1989).
%
\end{thebibliography}
\end{document}